\documentclass[11pt,a4paper]{article}
\usepackage{marvosym}
\usepackage{a4wide}
\usepackage{amsthm}
\usepackage{amsmath}
\usepackage{amssymb}
\usepackage{natbib}

\def\halmos{\hfill$\ \square$}
\def\myendproof{{}}
\def\fixorder#1{}

\def\marginpar#1{}
\def\endnote#1{}{}

\theoremstyle{plain}

\newtheorem{thm}{Theorem}
\newtheorem{prop}{Proposition}

\newtheorem{cor}{Corollary}

\theoremstyle{definition} 

\newtheorem{defn}{Definition}
\newtheorem{example}{Example}
\newtheorem{rmk}{Remark}

\def\boldrho{{\boldsymbol\rho}}
\def\boldsigma{{\boldsymbol\sigma}}

\def\boldone{{\mathbf 1}}

\def\ket#1{{ | #1 \rangle }}
\def\bra#1{{ \langle #1 | }}
\def\braket#1#2{{ \langle #1  |#2 \rangle }}
\def\ketbra#1#2{{\ket{#1}\bra{#2}}}
\def\braopket#1#2#3{{ \langle #1  |#2 | #3 \rangle }}

\title{On Quantum Statistical Inference}
\author{Ole E. Barndorff-Nielsen\\
MaPhySto
\thanks{MaPhySto is the Centre for Mathematical Physics and Stochastics,
funded by the Danish National Research Foundation, University of Aarhus, Denmark.}\\
\\ Richard D. Gill\\
Mathematical Institute, University of Utrecht and
\\
EURANDOM, Eindhoven, Netherlands\\
\\ Peter E. Jupp\\
School of Mathematics and Statistics, University of St Andrews, U.K.}

\date{July 26 , 2003}
     
\begin{document}
    
\maketitle

\begin{abstract}
Interest in problems of statistical inference connected to measurements of
quantum systems has recently increased substantially, in step with dramatic
new developments in experimental techniques for studying small quantum
systems. Furthermore, theoretical developments in the theory of quantum
measurements have brought the basic mathematical framework for the
probability calculations much closer to that of classical probability
theory. The present paper reviews this field and proposes and interrelates
a number of new concepts for an extension of classical statistical
inference to the quantum context.  (An earlier version of the paper containing
material on further topics is available as \texttt{quant-ph/0307189}).
\end{abstract}

\newpage 
\tableofcontents 
\newpage


\section{Introduction}\label{s:intro}

Quantum mechanics has replaced classical (Newtonian) mechanics as the 
basic paradigm for physics. From there it pervades chemistry, 
molecular biology, astronomy, cosmology,\dots\ . The theory is fundamentally
stochastic: the predictions of quantum mechanics are probabilistic.
When used to derive properties of matter, the stochastic nature of the
theory is typically swallowed up by the law of large numbers
(very large numbers, like Avogadro's, $10^{23}$). However, in some 
situations randomness does appear on the surface, most
familiarly in the random times of clicks of a Geiger-counter. Present-day
physicists, challenged by the fantastic theoretical promise of a
quantum computer, are carrying out experiments in which half a dozen
ions are held in an ion-trap and individually pushed into lower or
higher energy states, and into quantum superpositions 
of such joint states. The existence of 
these wavelike superpositions of combinations of distinct states 
of distinct objects is a fundamentally quantum 
phenomenon called \emph{entanglement}. Entanglement is of
enormous importance in quantum computation and quantum communication.
In other experiments, using supercooled
electric circuits, billions of electrons behave as a single quantum
particle which is brought into a wavelike superposition of macroscopically
distinct states (clockwise and anti-clockwise current flow, for instance).
This was recently achieved in Delft by \citet{mooijetal99} using a SQUID
(semiconducting quantum interference device). \citet{hannemannetal02}
recently implemented a Bayesian sequential adaptive design-and-estimation procedure
to determine the state of $12$ identically prepared two-level systems.

In these experiments, single quantum systems are individually
manipulated and probed. The outcomes of measurements are random,
with a probability distribution which depends on the one hand, on 
which \emph{quantum measurement} (the experimental design)
was carried out, and on the other hand, on the \emph{state}
of the quantum system being measured. If one does not know in advance
the state of the quantum system, or wants to use the measurement results
in order to prove that a certain state had been created, one is dealing
with statistical estimation and testing problems for data from a
probabilistic model with a rather elegant mathematical structure,
as we shall see.

By the nature of quantum mechanics, 
measurement of a quantum system disturbs the system. 
The complete specification of a particular experiment tells us not only how the
distribution of the data depends on the state of the quantum system being
measured, but also how the state of the system after the measurement
depends on its initial state and on the outcome which was observed.
This complete specification is described mathematically by a \emph{quantum instrument}.
Measuring the system in one way precludes measuring it 
simultaneously in a different way.
The total amount of information which can be obtained about an 
unknown parameter of the state of a quantum system is finite. 
Quantum physics delineates in a very precise way the class of all
possible instruments.
Thus, before looking at which experiments are practically feasible,
one can already investigate mathematically the limits of the
information which can be extracted from an unknown quantum system, leading to
advice on various experimental strategies.

The field of quantum statistical inference studies these problems in a
unified and systematic way. Established a quarter of a century ago in
the classic monographs of \citet{helstrom76} and \citet{holevo82}, it
is currently under vigorous renewed development, stimulated by 
experimental efforts in nanotechnology, and the rapid theoretical 
development of quantum communication, quantum cryptography, 
quantum computation, and quantum information theory.

Though real laboratory experiments involve highly complex models and
severe practical limitations, the basic theory and the basic statistical
issues should be accessible to a general statistical audience.
The most elementary models involve $2\times 2$ complex matrices,
some linear algebra and elementary probability. Such models already allow 
one to state problems of statistical design and inference which we are only just
starting to understand, and which are relevant to experimentalists and
theoreticians in quantum information.

The purpose of this paper is to introduce this problem area to the statistical 
community. We set up the basic statistical modelling in the simplest 
of settings, namely that of a small collection of
two-dimensional quantum systems. Depending on the context, such quantum
systems are called `spin-half systems' (the spin of an electron, for
instance), or `two-level atoms' (ground state and first excited state
for atoms in an ion trap, at very low temperature), or `qubits' 
(the `bits' of the RAM of a future quantum computer for which various
technologies are being currently explored; one possibility being a
supercooled aluminium ring in which an electric current might flow 
clockwise or anti-clockwise). Also covered is the polarisation of
photons, leading us to phenomena studied in quantum optics such as
violation of the \citet{bell64} inequalities in the \citet{aspectetal82b} experiment,
of great current interest; see \citet{weihsetal98}, \citet{gill01c}. Thus the same
mathematical and statistical modelling covers a multitude of applications.

\subsection{Overview}\label{ss:overview}

The paper is organised as follows. Section \ref{s:states} describes
the mathematical structure linking states of a quantum system,
possible measurements on that system, and the resulting state of
the system after measurement. Section \ref{s:lik} introduces quantum
statistical models and notions of quantum score and quantum
information, parallel to the score function and Fisher information
of classical parametric statistical models. In Section \ref{s:qeqtm} we 
introduce quantum exponential models and quantum transformation 
models, again forming a parallel with fundamental classes of
models in classical statistics. In Section \ref{s:exhaust} we describe
notions of quantum exhaustivity, quantum sufficiency and quantum cuts of 
a measurement, relating them to the classical notions of sufficiency
and ancillarity. We next turn,
in Section \ref{s:info},  to a study of the relation between 
quantum information and classical
Fisher information, in particular through Cram\'er--Rao type
information bounds. In Section \ref{s:qu-class} we discuss the interrelation 
between classical and quantum probability and statistics.
Finally, in Section \ref{s:other} we conclude with
remarks on further topics of potential interest to 
probabilists and statisticians. Sections \ref{s:qeqtm}, \ref{s:exhaust}, and
\ref{s:info} contain a considerable amount of new work.

This paper complements our more mathematical survey 
\citep*{barndorffnielsenetal01} on quantum statistical information.
A version of this paper with much further material (such as foundational 
questions, Bell inequalities, infinite dimensional spaces, continuous time 
observation of a quantum system) is available as  
\citet*{barndorffnielsenetal01b}.
Many further details can be found in \citet*{ barndorffnielsenetal03}.
\citet{gill01a} is a tutorial introduction to the basic modelling, while 
\citet{gill01b} is an introduction to large sample quantum estimation theory.
Some general references which we have found extremely useful are the 
books of \citet{isham95}, \citet{peres95}, \citet{gilmore94} and 
\citet{holevo82,holevo01book}. The reader is also referred to the
`bible of quantum information' \citet{nielsenchuang00}, which contains
excellent introductory material on the physics and the computer science,
and to the basic probability and statistics text \citet{williams01}
which recognises (Chapter 10) quantum probability as a topic which should be in every
statistician's basic education. Finally, the former Los Alamos National 
Laboratory preprint service for quantum physics, now at Cornell,
\texttt{quant-ph} at 
\texttt{http://arXiv.org} is an invaluable resource.

\section{States, Measurements and Instruments}\label{s:states}

\subsection{The Basics}

The state of a finite-dimensional quantum system 
is described or represented by a $d\times d$ matrix $\rho$ of complex
numbers, called the \emph{density matrix}. The number $d$ is the
dimension of the system and already the case $d=2$ is rich both in mathematical
structure and in applications, some of which were mentioned above.
We shall write $\mathcal H=\mathbb C^d$ for the Hilbert space of 
$d$-dimensional complex vectors, also called the state space of the system.
The inner product of vectors $\phi$ and $\psi$, written by physicists as
$\braket\phi\psi$ and by mathematicians as $\phi^*\psi$, equals 
$\sum\overline{\phi_i}\psi_i$  (the bar denotes complex conjugation). 
The length or norm of a vector is defined through 
$\|\phi\|^2=\braket\phi\phi$.

A density matrix $\rho$ has to be nonnegative and of
trace $1$, these properties being defined as follows.
The trace of a square matrix is defined in
the usual way as the sum of its diagonal elements.
The definition of nonnegativity is a little more complicated.
First, for an arbitrary complex matrix $X$ we define the adjoint $X^*$ of $X$ to be the
matrix obtained from $X$ by taking its transpose and replacing each 
element by its complex conjugate. 
An element $\psi$ of the state space $\mathcal H$ 
is to be thought of as a column vector and
hence $\psi^*$ is a row vector containing the complex conjugates of the
elements of $\psi$. Since $\rho$ is a $d\times d$ matrix, the
quadratic form $\psi^* \rho \psi$ is a complex scalar. 
The statement that $\rho$ is nonnegative means just that  $\psi^* \rho \psi$
is real and nonnegative for every $\psi\in\mathcal H$. 

Physicists would  write
$\ket\psi$ for the column vector $\psi$, $\bra\psi$ for the row vector $\psi^*$,
and $\braopket\psi\rho\psi$ for the number $\psi^* \rho \psi$. In particular,
$\braket\psi\psi=\|\psi\|^2$ is a number, while if $\|\psi\|=1$ then $\ketbra\psi\psi$
is the matrix which projects onto the one-dimensional subspace of $\mathcal H$ spanned
by $\psi$. This bra-ket notation, due to Dirac, appears at first sight merely to require
superfluous typing but it does gives a visual clue to the status of
various objects and moreover provides a short-hand whereby the name of the bra or ket $\psi$ 
can be replaced by some identifying words or symbols, as in $\ket{{\text{\Smiley}}}$, 
$\ket{{\text{\Frowny}}}$.

It can be shown that a nonnegative matrix is automatically self-adjoint, i.e.\ 
$\rho=\rho^*$. Self-adjoint matrices share some familar properties of symmetric real matrices: one can find an orthonormal basis of eigenvectors, and the eigenvalues are real numbers. The trace of a matrix equals the sum of the eigenvalues.
A nonnegative matrix has nonnegative eigenvalues. Thus the eigenvalues of $\rho$ can be interpreted as a probability distribution over $\{1,\dots,d\}$. As we shall see, this probability distribution has a
physical meaning: the state $\rho$ can be thought of as a probability mixture
over a collection of $d$ states, each associated with one of the eigenvectors 
and of a special type called a \emph{pure state}.
A probability mixture $\rho=\sum_i p_i \rho_i$
of density matrices is again a density matrix. This is the state obtained by
taking the quantum system in state $\rho_i$ with probability $p_i$.

\begin{example}[Pure state and mixed state]
Let $\ket{\phi_1}$, \dots, $\ket{\phi_d}$ denote an orthonormal basis of $\mathcal H$.  
If the basis is clear from the context, we can exploit the bra-ket 
notation and abbreviate these vectors to $\ket 1$, \dots, $\ket d$. 
Let $p_1$, \dots, $p_d$,
denote a probability distribution over $\{1,\dots,d\}$.  Write 
\begin{equation}\label{e:densmatrixdecomp}
\rho~=~\sum_i p_i \ketbra{i}{ i}.
\end{equation} 
Note that $\rho_i=\ketbra{i}{i}$ is a
$d\times d$ matrix of rank one. It represents the operator which projects 
an arbitrary vector into the one-dimensional subspace of $\mathcal H$ 
consisting of all (complex) multiples of $\ket{i}$. One can easily check that it
is a density matrix. Such a state, with density matrix being a rank-one projector
and characterised by a unit length \emph{state vector} in $\mathcal H$, is a \emph{pure state}.
It follows that $\rho$, a probability mixture of density matrices,
is also a density matrix. By the eigenvalue-eigenvector decomposition of
self-adjoint matrices, any density matrix can be written in the form (\ref{e:densmatrixdecomp}),
with the vectors $\ket{i}$ orthonormal.

If a density matrix $\rho$ is of rank $1$, one can write $\rho=\ketbra{\phi}{\phi}$ for some vector
$\ket{\phi}$ with $\|\phi\|^2=\braket{\phi}{\phi}=1$. The state is called a pure state and $\ket{\phi}$ is called
the state vector; it is unique up to a complex factor of modulus $1$. If the rank of
a density matrix is greater than $1$ then the state is called \emph{mixed}. It can be written
as a mixture of pure states in many different ways, especially if one does not insist
that the state vectors of the pure states are orthogonal to one another.
\halmos
\end{example}

The density matrix of a quantum system encapsulates in a very concise but
rather abstract way all the predictions one can make about future observations 
on that system, or more generally, all results of interaction of the system
with the real world. 

So far we have been using the word `measurement' in a rather loose way, but
at this point it is important to make the technical distinction 
between mathematical models for \emph{a measurement when we do not care 
about the state of the system after the measurement, but only about the
outcome}, and models for \emph{a measurement including the 
the state of the system after the measurement}.  The former is called a
\emph{measurement} and denoted by $M$; the latter, more complicated object, is called an
\emph{instrument} and denoted by $\mathcal N$.

Let us start with the simpler object, a measurement. 
Consider a measurement with discrete outcome, i.e.\  the sample space of the outcome is at most countable. From quantum theory it follows that any measurement whatsoever, i.e.\  any
experimental set-up, is described mathematically by a collection $M$
of  $d\times d$ matrices $m(x)$ indexed by the outcomes $x$ of the experiment.
The matrices have to be nonnegative (and hence also self-adjoint) and must add
up to the identity matrix $\mathbf 1$. Let us write $p(x;\rho,M)$ for
the probability that applying the measurement $M$ to the state 
$\rho$ produces the outcome $x$. Then we have the fundamental formula
\begin{equation}\label{e:measprob}
p(x;\rho,M)~=~\mathrm{trace}(\rho m(x)).
\end{equation}

One can see that this expression indeed defines a bona-fide probability
distribution as follows. Writing $\rho=\sum p_i\ketbra{\phi_i}{\phi_i}$
and permuting cyclicly the elements in a trace of a product of matrices,
one finds $\mathrm{trace}(\rho m(x))=\sum p_i \mathrm{trace}( \ketbra{\phi_i}{\phi_i} m(x))
=\sum_i p_i \braopket {\phi_i} {m(x)} {\phi_i}$.
Thus, since $m(x)$ is a nonnegative matrix and the $p_i$ are probabilities,
$p(x;\rho,M)$ is a nonnegative real number. Moreover, the sum
over $x$ of these numbers is
$\sum_x \mathrm{trace}(\rho m(x)) =  \mathrm{trace}(\rho \sum_x m(x))
= \mathrm{trace}(\rho\boldone)= \mathrm{trace}(\rho)=1$.

A \emph{quantum statistical model} is a model for a partly or completely unknown state.
This means that the state $\rho$ is supposed to depend on an unknown 
parameter $\theta$ in some parameter space $\Theta$. 
Write $\boldrho=(\rho(\theta):\theta\in\Theta)$.
When we apply a measurement $M$ to a quantum system from this model,
the outcome has probability density
\begin{equation}\label{e:pxtheta}
p(x;\theta,M)~=~\mathrm{trace}(\rho(\theta) m(x)).
\end{equation}
Thus given the measurement and the quantum statistical model, a classical 
statistical inference problem is defined. Very important problems also arise when 
the measurement itself is indexed by an unknown parameter, 
but for reasons of space 
we do not address these here.

In principle, any measurement $M$ whatever could be implemented as
a laboratory experiment. Equation (\ref{e:pxtheta}) tells us implicitly how much
information about $\theta$ can be obtained from a given experimental set-up
$M$. One may try to choose $M$ in such a way as to maximise
the information which the experiment will give about $\theta$. Such experimental design
problems are a main subject of this paper.

Often we are interested also in the state of the system after the measurement.
In this case we need the more general notion of \emph{instrument}. An instrument 
$\mathcal N$ (more precisely, a `completely positive instrument')
is represented by a family of collections of $d\times d$ matrices
$n_i(x)$ satisfying $\sum_x\sum_i n_i(x)^* n_i(x)=\mathbf 1$ but being 
otherwise completely arbitrary. The index $x$ refers to the observed outcome
of the measurement, the index $i$ could be thought of as `missing data'.
Define $m(x)=\sum_i n_i(x)^* n_i(x)$. It 
follows that the matrices $m(x)$ are nonnegative (and self-adjoint) and add to
the identity matrix, and thus represent a measurement (in the narrow or
technical sense) $M$.
When we apply the instrument $\mathcal N$ to the
quantum system, the outcome has the same probability density as
(\ref{e:measprob}), but we write it out in terms of $\mathcal N$ as
\begin{equation}\label{e:instrp}
p(x;\rho,\mathcal N)~=~ \sum_i \mathrm{trace}(\rho n_i(x)^* n_i(x))
\end{equation}
and the state of the system after applying the measurement, conditioned
on observing the outcome $x$, is
\begin{equation}\label{e:instrsigma}
\sigma(x;\rho,\mathcal N)~=~\frac {\sum_i n_i(x) \rho n_i(x)^* }
                    {   \sum_i \mathrm{trace}(\rho n_i(x)^* n_i(x)) }.
\end{equation}
The reader should check that the expression for $\sigma(x;\rho,\mathcal N)$
does define a bona-fide density matrix (nonnegative, trace 1). In some important
practical problems the instrument itself depends on an unknown
parameter, but here we suppose it is completely known.

It follows from quantum physics that whatever one can do to
a quantum system has to have the form of a quantum instrument. Moreover,
in principle, any quantum instrument whatsoever could be realised
by some experimental set-up. Usually in the theory one starts by postulating
some natural physical properties of the transformation from \emph{input} or \emph{prior state} to
\emph{output} or \emph{posterior state} and data, and derives (\ref{e:instrp}) 
and (\ref{e:instrsigma}), which are then called
the \emph{Kraus representation} of the instrument, as a theorem. Here it is more
convenient to start with (\ref{e:instrp}) and (\ref{e:instrsigma}). Further discussion
and references are given in section \ref{ss:instru}.

One could consider applying two different quantum instruments, one after
the other, to the same quantum system. One might even allow the choice of
second instrument to depend on the outcome obtained from the first.
The composition of two instruments in this way defines a new one; it
is not difficult to express the matrices $n_i(x)$ of the new instrument in terms
of those of the old ones.
Another way to get new instruments from old is by coarsening. Suppose one
applies one instrument to a quantum system, then applies a many-to-one function
of the outcome, and discards the original data. The new instrument can be
written down in terms of the old by relabelling the matrices $n_i(x)$ with new
index $j$ and variable $y$  in obvious fashion.

In classical statistics, central notions such as sufficiency are connected to
decomposing statistical models into parts (marginal and conditional distributions),
and to reducing statistical models by reducing data.
Starting with a quantum statistical model with density matrices $\rho$
depending on a parameter $\theta$, possibly with nuisance parameters too,
it is now natural to ask whether notions akin to sufficiency and ancillarity can be 
developed for instruments. For instance, it might happen that the posterior state 
of a quantum system after applying a certain instrument
no longer depends on the unknown parameter.

In the next subsection we shall work out many of these notions for the important special case
of a two-dimensional quantum system. But first we present two special examples,
connecting the notion of instrument to the classical notions (in quantum physics)
of observables and unitary transformations.

\begin{example}[Simple instruments, simple measurements]
Let $x_1$, \dots, $x_d$ denote $d$ distinct real numbers and let
$\ket{\psi_x}$, $x\in\{x_1\dots,x_d\}$ be an orthonormal basis of $\mathcal H$
indexed by the numbers in $\mathcal X=\{x_1,\dots,x_d\}$.
We can now define an instrument $\mathcal N$ with outcomes in $\mathcal X$
by supposing that the index $i$ takes only one value, let us call it $0$,
and taking $n_0(x)=\ketbra{\psi_x}{\psi_x}$.
This matrix is self-adjoint and idempotent (equal to
its square). Therefore the corresponding matrices $m(x)$ are given by
$m(x)=\ketbra{\psi_x}{\psi_x}$ too, and they sum to the identity matrix: the sum of
projectors onto orthogonal one-dimensional subspaces spanning the whole
space, is the identity. This shows that $\mathcal N$ is indeed an instrument,
though of very special form indeed.

We can now compute the probability of observing the outcome $x$ and the
posterior state of the quantum system, given the outcome is $x$, when the
quantum system is originally in the state $\rho=\sum_i p_i\ketbra{\phi_i}{\phi_i}$.
A straightforward calculation shows that they are given as follows:
\begin{equation}
p(x;\rho,\mathcal N)~=~ \braopket{\psi_x}\rho{\psi_x}~=~\sum_i p_i  | \braket{\psi_x}{ \phi_i}  |^2
\end{equation}
\begin{equation}
\sigma(x;\rho,\mathcal N)~=~ \ketbra{\psi_x}{\psi_x}.
\end{equation}
These formulae can be interpreted probabilistically as follows.
The quantum system was initially in the pure state with density
matrix $\ketbra{\phi_i}{\phi_i}$ with probability $p_i$. On being measured with
the instrument $\mathcal N$, the system jumped to the pure state
with density matrix  $\ketbra{\psi_x}{\psi_x}$  producing the outcome $x$,
with probability $| \braket{\psi_x}{\phi_i}|^2$.

Let $X=\sum_x x \ketbra{\psi_x}{\psi_x}$. This is a self-adjoint matrix with eigenvalues
$x_1,\dots,x_d$ and eigenvectors $\ket{\psi_1},\dots,\ket{\psi_d}$. One says that the
instrument $\mathcal N$ corresponds to the \emph{observable} $X$. 
`Measuring the observable' with this instrument produces one of the eigenvalues, 
and forces the system into the corresponding eigenstate. 
If the quantum system starts in a pure state with state vector $\phi$, 
then it jumps to the eigenstate $\psi_x$ with probability
$ |\braket{ \psi_x}{\phi } |^2$.

Suppose now $X$ is an arbitrary self-adjoint matrix. Let $\mathcal X=\{x_1,\dots,x_{d'}\}$
denote its \emph{distinct} eigenvalues. Let $\Pi(x)$ denote the matrix which projects
onto the eigenspace corresponding to eigenvalue $x$, not necessarily one-dimensional.
Thus $X=\sum_x x \Pi(x)$.
Define $n_0(x)=m(x)=\Pi(x)$.
We see again that the matrices $n_0(x)$ define an instrument $\mathcal N$,
and the matrices $m(x)$ define a corresponding measurement $M$.
When this instrument is applied to the quantum system $\rho=\sum_i p_i\ketbra{\phi_i}{\phi_i}$,
one obtains the outcome $x$ with probability $\sum_i p_i\| \Pi(x) \ket{\phi_i }\|^2$.
One may compute that the final state is the mixture, according to the posterior probabilities
that the initial state was $\ket{\phi_i}$ given that the outcome is $x$, of the pure states with
state vectors equal to the normalised projections $\Pi(x)\ket{\phi_i}/\| \Pi(x) \ket{\phi_i }\|$.
Yet again we have the probabilistic interpretation, that with probability $p_i$ the 
quantum system started in the pure state with state vector $\ket{\phi_i}$.
On measuring the observable $X$, the state vector is projected into one of the eigenspaces,
with probabilities equal to the squared lengths of the projections. One gets to observe the
corresponding eigenvalue. The posterior state is the mixture of these different pure states
according to the posterior distribution of initial state given data $x$.

When one measures the observable $X=\sum_x x \Pi(x)$ with the corresponding simple
instrument or simple measurement, the probability of each eigenvalue $x$ is
$\mathrm{trace}(\rho\Pi(x))$. It follows that the expected value of the outcome
is $\mathrm{trace}(\rho X)$.  More generally, let $f$ be some real function.
One may define the function $f$ of the observable $X$ by $Y=\sum_xf(x)\Pi(x)$.
This is the self-adjoint matrix with the same eigenspaces, and with eigenvalues 
equal to the function $f$ of the eigenvalues of $X$. If the function $f$ is many-to-one
then some eigenspaces may have merged---consider the function `square' for instance.
It follows that the expected value of the function $f$ of the outcome of measuring $X$
is given by the elegant formula $\mathrm{trace}(\rho f(X))$.
We call this rule \emph{the law of the unconscious quantum physicist} 
since it is analogous to the law of the unconscious statistician, according 
to which the expectation of a function $Y=f(X)$ of a random variable $X$
may be calculated by an integration (i) over the underlying probability 
space, (ii) over the outcome space of  $X$, (iii) over the outcome space 
of $Y$. Note however that the simple instruments corresponding to $X$ and to $Y$ are 
different, and moreover neither is equal to the instrument `measure $X$, 
but record only $y=f(x)$'.

This calculus of expected values of (outcomes of measuring) observables is the basis of the
mathematical theory called \emph{quantum probability}; for some further remarks
on this see Section \ref{s:qu-class}.

Two observables $P$, $Q$ are called \emph{compatible} if as operators 
they commute: $PQ=QP$. A celebrated
result of von Neumann is that
observables $Q$ and $P$ are compatible if and only if they are both 
functions $f(R)$, $g(R)$ of a third observable $R$. 
Taking $R$ to have as coarse a 
collection of eigenspaces as possible, one can show that the results 
of the following three instruments are identical: 
the simple instrument for $Q$ followed by the simple instrument for $P$,
recording the values $q$ of $Q$ and $p$ of $P$; the simple instrument 
for $P$ followed by the simple instrument for $Q$, recording the values 
$q$ of $Q$ and $p$ of $P$; and the simple instrument for $R$, recording the
values $q=f(r)$ and $p=g(r)$ where $r$ is the observed value of $R$.
It follows that the probability distribution of the outcome of 
measurement of an observable $P$ is not altered when it is measured
(simply, jointly) together with any other compatible observables.

An instrument such that the 
index $i$ takes only one value, say $0$, and such
that all $n_0(x)$ are projectors onto orthogonal subspaces of $\mathcal H$, together
spanning the whole space, is called a \emph{simple instrument}. The corresponding measurement
is called a \emph{simple measurement}. Simple instruments and measurements stand in one-to-one
correspondence with observables. The rule for the transformation of the state under a
simple instrument is called the \emph{L\"uders-von Neumann projection postulate}.
\halmos
\end{example}

\begin{example}[Instrument with no data]
It is possible that the quantum instrument $\mathcal N$ transforms the quantum
system $\rho$ without actually producing any outcome $x$: in the definition of
an instrument, simply take the outcome space to consist of a single element,
let us call it $0$.
Then the state $\rho$ is transformed by the instrument into the state
$\sum_i n_i(0) \rho n_i(0)^*$ where the $n_i(0)$ are matrices
satisfying $\sum_i  n_i(0)^*n_i(0)=\mathbf 1$. A very special case is
obtained when there is also only one value of the index $i$ and the
instrument is defined by a single matrix $U=n_0(0)$ satisfying $U^*U=UU^*=\mathbf 1$.
State $\rho$ is transformed into $U\rho U^*$. Such a matrix $U$ is called
\emph{unitary} and it corresponds to an orthogonal change of basis. 
A unitary transformation is invertible, and corresponds to the reversible
time evolution
of an isolated quantum system; see Subsection~\ref{ss:schroed} below.
\halmos
\end{example}

\subsection{Spin-half}\label{ss:spin12}

Our examples will concern mainly the spin of spin-half particles, 
where the dimension $d$ of $\mathcal{H}$ is $2$. Unfortunately,
it would take us too far afield to explain the significance of the word \emph{half}.
The classic example in this context is the 1922 experiment of Stern and 
Gerlach, see \citet[{\ }Section 1.4]{brantdahmen95}, to determine the
size of the magnetic moment of the electron. The electron was conceived
of as spinning around an axis and therefore behaving as a magnet
pointing in some direction. Mathematically, each electron carries a
vector `magnetic moment'. One might expect the sizes of the magnetic
moment of all electrons to be the same, but the directions to be
uniformly distributed in space. Stern and Gerlach made a beam of silver
atoms move transversely through a steeply increasing 
vertical magnetic field. A silver atom has 47 electrons 
but it appears that the magnetic
moments of the 46 inner electrons cancel and essentially only one
electron determines the spin of the whole system. Classical physical
reasoning predicts that the beam would emerge spread out vertically
according to the component of the spin of each atom (or
electron) in the direction of the gradient of the magnetic field. 
The spin itself would not be altered by passage through the magnet.
However, amazingly, the emerging beam consisted of just two well
separated components, as if the component of the spin vector in the
vertical direction of each electron could take on only two different
values, which in fact are $\pm\frac12$ in appropriate units.

This example fits into the following mathematical framework.
Take $d=2$, then $\mathcal{H}=\mathbb C^{2}$ 
and $\rho$ is a $2\times2$ matrix
$$
\rho~=~\left(
\begin{matrix}
\rho_{11} & \rho_{12}\\
\rho_{21} & \rho_{22}
\end{matrix}
\right)
$$
with $\rho_{21}=\overline\rho_{12}$ and
$\rho_{11}$ and $\rho_{22}$ real and nonnegative and adding to $1$.
The matrix has non-negative real eigen\-values $p_{1}$ and $p_{2}$ 
also adding to $1$.

In this case the density matrices of pure states can be
put into one-to-one correspondence with the unit sphere $S^2$,
the surface of the unit ball in
real, $3$-dimensional space. Directions in the sphere correspond to directions of
spin. This geometric representation is known in 
theoretical physics as the \textit{Poincar\'{e} sphere}, 
in quantum optics as the \textit{Bloch sphere}, 
and in complex analysis as the \textit{Riemann sphere}.
The mixed states, i.e.\  the convex combinations of pure states, correspond to points in
the interior of the ball. The mapping from states (matrices) to points in the unit ball
is affine, as we shall now show.

Any real linear combination of self-adjoint
matrices is again self-adjoint. Since the $2$ diagonal elements of a self-adjoint
matrix must be real, and the $2$ off-diagonal elements are one another's complex 
conjugate, just $4$ real parameters are needed to specify any such matrix.
By inspection one discovers that the space of self-adjoint matrices is spanned 
by the identity matrix
$$
\boldone =\sigma_{0}=\left(
\begin{matrix}
1 & 0\\
0 & 1
\end{matrix}
\right)  
\thinspace ,
$$
together with the three \emph{Pauli matrices}
$$
\sigma_{x}=\left(
\begin{matrix}
0 & 1\\
1 & 0
\end{matrix}
\right)  
\quad
\sigma_{y}=\left(
\begin{matrix}
0 & -i \\
i & 0
\end{matrix}
\right)  
\quad
\sigma_{z}=\left(
\begin{matrix}
1 & 0\\
0 & -1
\end{matrix}
\right)  
\thinspace .
$$
Note that $\sigma_{x},\sigma_{y}$ and $\sigma_{z}$ satisfy the
\emph{commutation relations}
\begin{align*}
\lbrack \sigma_{x},\sigma_{y}] ~&=~ 2i\sigma_{z}  \\
\lbrack \sigma_{y},\sigma_{z}] ~&=~ 2i\sigma_{x}  \\
\lbrack \sigma_{z},\sigma_{x}] ~&=~ 2i\sigma_{y} ,
\end{align*}
where, for any operators $A$ and $B$, their commutator $[A,B]$ 
is defined as $AB - BA$. Note also that
$$
\sigma_{x}^{2}=\sigma_{y}^{2}=\sigma_{z}^{2}=\boldone 
\thinspace .
$$

Any pure state has the form $\ketbra\psi\psi$ for some 
unit vector $|\psi\rangle$ in $\mathbb C ^{2}$. 
Up to a complex factor of modulus $1$ (the \emph{phase}, which does
not influence the state), we can write $\ket\psi$ as
$$
\ket\psi =
\left( 
\begin{matrix}
e^{-i\phi/2}\cos(\theta/2) \\
e^{i\phi/2}\sin (\theta/2)
\end{matrix}
\right)
\thinspace .
$$
The corresponding pure state is
$$
\rho=\left(
\begin{matrix}
\cos^{2}(\theta/2) & e^{- i\phi}\cos(\theta/2)\sin(\theta/2)\\
e^{i\phi}\cos(\theta/2)\sin(\theta/2) & 
\sin^{2}(\theta/2)
\end{matrix}
\right)  
\thinspace .
$$
A little algebra shows that $\rho$ can be written as 
$\rho=(\boldone +u_{x}\sigma
_{x}+u_{y}\sigma_{y}+u_{z}\sigma_{z})/2=
\frac{1}{2}(\boldone +\vec{u}\cdot\vec{\sigma})$, 
where $\vec{\sigma}=(\sigma_{x},\sigma_{y},\sigma_{z})$ are the three 
Pauli spin matrices and $\vec{u}=(u_{x},u_{y},u_{z})=\vec{u}(\theta,\phi)$ 
is the point on the unit sphere with polar coordinates $(\theta,\phi)$.

An arbitrary mixed state is obtained by averaging pure states 
$\rho=\frac12(\mathbf 1+\vec u\cdot\vec\sigma)$
with respect to any probability distribution over real unit vectors $\vec u$. The
result is a density matrix of the form $\rho=\frac12(\mathbf 1+\vec a\cdot\vec\sigma)$,
where $\vec a$ is the centre of mass (a point in the unit ball) of the distribution of pure states
seen as a distribution over the unit sphere. The coordinates of $\vec a$ are called the \emph{Stokes 
parameters} when we are using this model to describe polarization of a photon, rather than
spin of an electron.

\subsection{Superposition and Mixing}\label{ss:super}

Given two state vectors $\ket{\phi_1}$ and $\ket{\phi_2}$ and two complex numbers $c_1$, $c_2$,
the state vector $(c_1\ket{\phi_1}+c_2\ket{\phi_2})/\|c_1\ket{\phi_1}+c_2\ket{\phi_2}\|$ is called
the \emph{quantum superposition} of the original two states, with complex weights $c_1$, $c_2$.
This is a completely different way of combining two states from the mixture $p_1\ketbra{\phi_1}{\phi_1}+
p_2\ketbra{\phi_2}{\phi_2}$. (Sometimes the latter is called an `uncoherent mixture' and the
former a `coherent mixture'.) For example, consider the case $d=2$, let $\ket{\phi_1}$ and $\ket{\phi_2}$
form an orthornormal basis of $\mathcal H=\mathbb C^2$, and suppose $c_1=c_2=1/\sqrt 2$, $p_1=p_2=1/2$.
We consider the equal weights superposition and the equal weights mixture of the pure states 
 $\ket{\phi_1}$ and $\ket{\phi_2}$, showing how some measurements are able to distinguish between
these states, whereas others do not.

The two matrices $m(1)=\ketbra{\phi_1}{\phi_1}$, $m(2)=\ketbra{\phi_2}{\phi_2}$ define a measurement
$M^\phi$ with two possible outcomes $1$ and $2$, say. The probability distributions of the outcome under the
superposition and under the mixture just defined are identical (probabilities $1/2$ for each of the outcomes $1$
and $2$). 

Define now a new orthonormal basis
$\ket{\psi_1}=(\ket{\phi_1}+\ket{\phi_2})/\sqrt 2$, $\ket{\psi_1}=(\ket{\phi_1}-\ket{\phi_2})/\sqrt 2$.
Corresponding to this basis, one can construct a measurement $M^\psi$ in the same way as
before. When the superposition is measured with $M^\psi$, the outcome $1$ is certain and
the outcome $2$ is impossible. However, when the mixture is measured with $M^\psi$, the two outcomes
have equal probability $1/2$.

It is very important to note that a pure state can be expressed as a superposition of others, and
a mixed state as a mixture of others, in many different ways.

\subsection{The Schr\"{o}dinger Equation}\label{ss:schroed}

Typically the state of a quantum system
undergoes an evolution with time under the influence of an external field. The
most basic type of evolution is that of an arbitrary initial state $\rho_{0}$
under the influence of a field with Hamiltonian $H$. This takes the form
$$
\rho_{t}=e^{tH/i\hbar}\rho_{0}e^{-tH/i\hbar}
\thinspace ,
$$
where $\rho_{t}$ denotes the state at time $t$, $\hbar=1.05\times10^{-34}$ J sec is Planck's constant,
and $H$ is a self-adjoint operator on $\mathcal{H}$. If $\rho_{0}$ is a pure
state then $\rho_{t}$ is pure for all $t$ and we can choose unit vectors
$\psi_{t}$ such that $\rho_{t}=|\psi_{t}\rangle\langle\psi_{t}|$ and
\begin{equation}
\psi_{t}=e^{tH/i\hbar}\psi _0
\thinspace .\label{evoln}
\end{equation}
Equation (\ref{evoln}) is a solution of the celebrated 
\emph{Schr\"{o}dinger equation} $i\hbar(\mathrm d/\mathrm d t)\psi=H\psi$
or equivalently $i\hbar(\mathrm d/\mathrm d t)\rho=[H,\rho]$.
The matrix $e^{tH/i\hbar}$ is unitary. Conversely, every unitary matrix
$U$ can be written in the form $e^{tH/i\hbar}$ for some self-adjoint matrix $H$
and some time  $t$ and hence can be obtained by looking at some Schr\"odinger
evolution at a suitable time.

\subsection{Separability and Entanglement}\label{ss:entang}

When we study several quantum systems (with Hilbert spaces
$\mathcal{H}_1$, \dots, $\mathcal{H}_m$) interacting together, 
the natural model for the combined system has as its Hilbert space the 
tensor product $\mathcal{H}_1 \otimes \dots \otimes \mathcal{H}_m$.
Then a state such as $\rho_1 \otimes \dots \otimes \rho_m$ represents
`particle 1 in state $\rho_1$ and \dots and particle $m$ in state $\rho_m$'.
Suppose the states $\rho_{i}$ are pure with state vectors 
$\ket{\psi_{i}}$. Then the product state we have just defined is also
pure with state vector $\ket{\psi_1} \otimes \dots \otimes \ket{\psi_m}$.
A mixture of such states is called \emph{separable}. 

On the other hand,
according to the superposition principle, a complex superposition of
such state vectors is also a possible state vector of the interacting
systems.  Pure states whose state vectors  cannot be written in the product form 
$\ket{\psi_1} \otimes \dots \otimes \ket{\psi_m}$ are called \emph{entangled}.
The same term is used for mixed states which cannot be written 
as a mixture of pure product states. A state which is not entangled
is separable.
The existence of entangled states is respons\-ible for 
extraordinary quantum phenomena,
which scientists are only just starting to harness (in quantum
communication, comput\-ation, teleportation, etc.). 

An important physical feature of unitary evolution in a tensor 
pro\-duct space is that, in general, it does not preserve 
separability of states. 
Suppose that the state $\rho _1 \otimes \rho _2$ evolves 
according to the Schr\"{o}dinger operator $U_{t}=e^{tH/i\hbar}$ 
on $\mathcal{H}_{1}\otimes \mathcal{H}_{2}$. 
In general, if $H$ does not have the special form 
$H_1 \otimes \boldone  _2 +  \boldone  _1 \otimes H_2$,
the corresponding state at any non-zero time is 
entangled. The notorious \emph{Schr\"{o}dinger Cat}, is 
a consequence of this phenomenon of entanglement.
For an illustrative discussion of this see, 
for instance, \citet[{\ }Sect.\ 8.4.2]{isham95}. 

Consider a product quantum system with density matrix  $\rho$.
On its own, the first component has \emph{reduced density matrix}
$\rho_1$ obtained by ``tracing out'' the second component,
$(\rho_1)_{ij}=\sum_k(\rho)_{ik,jk}$. This procedure corresponds to
computing a marginal from a joint probability distribution. 
Any mixed state can be considered as the 
reduction to the system of interest of a pure state on an enlarged,
joint system. For instance, the completely mixed state $\mathbf 1/d$
is the result of tracing out the second component from the pure state
$\sum_j\ket j\otimes\ket j/\sqrt d$ on the product space formed from two
copies of the original space.

\subsection{Further Theory of Measurements}\label{ss:meas}

\begin{example}[Spin-half, cont.]\label{e:spin12c1}
For any unit vector $\psi$ of $\mathbb C ^{2}$, let $\psi^\perp$
denote the unit vector orthogonal to it (unique up to a complex phase).
The observable 
(self-adjoint matrix)
$2\ketbra\psi\psi-\boldone 
= \ketbra\psi\psi-\ketbra{\psi^{\perp}}{\psi^{\perp}}$ defines a simple
instrument.
It has eigenvalues $1$ and 
$-1$ and one-dimensional eigenspaces spanned by $\psi$ and 
$\psi^{\perp}$.
This observable corresponds to the spin of the particle in the direction 
(on the Poincar\'e sphere) defined by $\psi$. When `the spin is measured
in this direction' meaning, when this observable is measured,
 the result (in appropriate units) is either $+1$ or $-1$.
Moreover, after the measurement has been carried out, the particle
is in the pure state of spin in the corresponding direction.
We mentioned two such measurements in Section \ref{ss:super} 
on mixing and superposition.

In particular, with outcome space $\mathcal{X}=\{-1,1\}$, the specific\-ation
\begin{align*}
n_0(+1)~=~m(+1) ~&=~ {\textstyle{\frac12}}(\boldone +\sigma_{x})\\
n_0(-1)~=m(-1)  ~&=~ {\textstyle{\frac12}}(\boldone -\sigma_{x})
\end{align*}
defines a simple instrument (where the index takes only  one value). 
It corresponds to the observable $\sigma_{x}$: spin in the $x$-direction.
\halmos\end{example}

We next discuss the notion of \emph{quantum 
random\-isation},
whereby adding an auxiliary quantum system to a system 
under study gives one further possibilities for probing the system of 
interest. This also connects to the important notion of \emph{realisation}, i.e.\ 
representing a measurement by a simple measurement on a 
quantum randomised extension.

Suppose given a Hilbert space $\mathcal{H}$, and a pair 
$(\mathcal{K},\rho_{a})$, where $\mathcal{K}$ is a Hilbert space and 
$\rho_{a}$ is a state on $\mathcal{K}$. 
Any measurement $\widetilde{M}$ on the product space $\mathcal{H}\otimes\mathcal{K}$
induces a measurement $M$ on $\mathcal{H}$ by the defining relation
\begin{equation}
\mathrm{trace}\left(\rho m(x) \right) = 
\mathrm{trace}\left( (\rho\otimes\rho_{a}) {\widetilde{m}}(x)\right)  
\quad\mbox{for all states $\rho$ on $\mathcal{H}$, all outcomes $x$}
\thinspace . \label{ancilla}
\end{equation}
The pair $(\mathcal{K},\rho_{a})$ is called an \emph{ancilla}. The following
theorem (Holevo's extension of Naimark's Theorem) states that any
measurement $M$ on $\mathcal H$ has the form
(\ref{ancilla}) for some ancilla $(\mathcal{K},\rho_{a})$ and some 
simple measurement  ${\widetilde{M}}$ on
$\mathcal{H}\otimes
\mathcal{K}$. The triple $(\mathcal{K},\rho_{a},{\widetilde{M}})$ is called 
a \emph{realisation} of $M$ (the words \emph{extension} or 
\emph{dilation} are also used sometimes). 
Adding an ancilla before taking a simple measurement 
could be thought of as \emph{quantum randomisation}. 

\begin{thm}[\citealt{holevo82}]\label{t:holevo}
For every measurement $M$ on $\mathcal{H}$, there is an ancilla 
$(\mathcal{K},\rho_{a})$ and a simple measurement ${\widetilde{M}}$ on
$\mathcal{H} \otimes\mathcal{K}$ which form a 
realisation of $M$.
\end{thm}

We use the term `quantum randomisation' because of its analogy with the mathematical 
representation of randomisation in classical statistics, whereby one 
replaces the original probability space with a product space,
one of whose components is the original space of 
interest, while the other corresponds to an independent random 
experiment with probabilities under the control of the experimenter.
Just as randomisation in classical statistics is sometimes needed to 
solve optimisation problems of statistical decision theory, 
so quantum randomisation sometimes allows for strictly better solutions 
than can be obtained without it.

Here is a simple spin-half example of a non-simple measurement which cannot be 
represented without \emph{quantum} randomisation. 

\begin{example}[The triad]
The triad, or Mercedes-Benz logo, has an outcome space consisting of just three 
outcomes:
let us call them $1$, $2$ and $3$. Let $\vec v_{i}$, $i=1,2,3$, denote three 
unit vectors in the same plane through the origin in $\mathbb R^{3}$, 
at angles of $120^{\circ}$ to one another. Then the matrices 
$m(i)=\frac 13 (\mathbf 1 +\vec 
v_{i}\cdot\vec\sigma)$  define a (non-simple) measurement $M$ 
on the sample space $\{1,2,3\}$. It turns up as the optimal solution to 
the decision problem: suppose a spin-half system is generated in one of the 
three states $\rho_{i}=\frac 12 (\mathbf 1 -\vec 
v_{i}\cdot\vec\sigma)$, $i=1,2,3$, with equal probabilities. 
What decision rule gives the maximum probability of guessing the 
actual state correctly? There is no way to equal the success 
probability of $M$ if one restricts attention to simple measurements, 
or to classically randomised procedures based on simple measurements.
\halmos\end{example}

Finally, we introduce some further terminology concerning measurements.
Given a measurement $M$ and a function $T$ from its outcome space 
$\mathcal X$
to another space $\mathcal Y$, one can define a new measurement 
$M'=M\circ T^{{-1}}$ with outcome space $\mathcal Y$. It corresponds to 
restricting attention to the function $T$ of the outcome of the first measurement 
$M$. We call it a \emph{coarsening} of the original measurement, and 
we say that $M$ is a \emph{refinement} of $M'$.

So far we have restricted attention to measurements with discrete outcome space.
More generally, one considers measurements with outcomes in an arbitrary
measure space $(\mathcal X,\mathcal A)$ where $\mathcal A$ is a sigma-algebra
of measurable subsets of $\mathcal X$. Such measurements are defined by a collection
of matrices $M(A)$ which are nonnegative, sigma-additive over $\mathcal A$,
and such that $M(\mathcal X)=\mathbf 1$. The probability that the outcome lies
in the set $A\in\mathcal A$ is $ \mathrm{trace}(\rho M(A))$.
A measurement $M$ is called \emph{dominated} by a (real, sigma-finite) 
measure $\nu$ on the outcome space, if there exists a non-negative 
self-adjoint matrix-valued function $m(x)$, called the density of $M$, such that 
$M(A)=\int_{A}m(x)\nu(\mathrm d x)$ for all $B$. When $\mathcal H$ is
finite dimensional, as in this paper, 
every measurement is dominated: take $\nu(A)=\mathrm{trace}(M(A))$.
In the dominated case, the outcome of the measurement has a probability
distribution with density $p(x;\rho,M)=\mathrm{trace}(\rho m(x))$
with respect to $\nu$. If the outcome space is discrete and $\nu$ is counting
measure, then these notations are linked to our original setup by $m(x)=M(\{x\})$,
$M(A)=\sum_{x\in A} m(x)$.

To exemplify these notions, suppose for some dominated measurement $M$ one can write 
$m(x)=m_{1}(x)+m_{2}(x)$ for two non-negative self-adjoint 
matrix-valued functions $m_{1}$ and $m_{2}$.
Define $M'$ to be the measurement on the outcome space 
$\mathcal X'=\mathcal X\times\{1,2\}$ with density $m(x,i)=m_{i}(x)$, $(x,i)\in 
\mathcal X'$ with respect to the 
product of $\nu$ with counting measure. Then $M'$ is a refinement of $M$ .

We described earlier how one can form product spaces from separate 
quantum systems, leading to notions of product states, separable 
states, and entangled states. Given a measurement $M$ on one component of a 
product space, one can naturally talk about `the same measurement' on the 
product system. It has components $M(A)\otimes\boldone$. Given 
measurements $M$ and $M'$ defined on the two components of a product 
system, one can define in a natural way the measurement `apply 
$M$ and $M'$ simultaneously to the first and second component, respectively': 
its outcome space is the product of the two outcome spaces, and it is defined 
using obvious notation by $(M\otimes M')(A\times A')=M(A)\otimes M'(A')$.

A measurement $M$ on a product space is called \emph{separable} if it 
has a density $m$ such that each $m(x)$ can be written as a positive 
linear combination of tensor products of non-negative components. It can 
then be thought of as a coarsening of a measurement with density $m'$
such that each $m'(y)$ is a product of non-negative 
components.

\subsection{Further Theory of Instruments}\label{ss:instru}

Just as we want to allow measurements also to take on continuous values,
so we need instruments to do the same.

Consider an instrument $\mathcal N$ with outcomes $x$ in the measurable space
$(\mathcal X, \mathcal A)$. Let $\pi(\mathrm d x;\rho,\mathcal N)$ denote the
probability distribution of the outcome of the measurement, and let
$\sigma(x;\rho,\mathcal N)$ denote the posterior state when the prior 
state is $\rho$ and the outcome of the measurement is $x$. It follows from the
laws of quantum mechanics that the only physically feasible instruments 
have a special form, generalising in a natural way the definitions we gave for
the discrete case. Namely, corresponding to 
$\mathcal N$ there must exist a $\sigma$-finite measure 
$\nu$ on $\mathcal{X}$ (which without loss of generality, can be taken to be a probability 
measure) and a collection of matrix-valued measurable functions $n_i$ of $x$
indexed by a finite or countable index $i$, such that
$$
\sum_{i} \int _{\mathcal{X}} n_i(x)^*  n_i(x) \nu (\mathrm d x) = 
\boldone 
\thinspace ;
$$
the posterior states for $\mathcal N$ are given by
\begin{equation}\label{e:instrsigmagen}
\sigma(x;\rho,\mathcal N) = \frac{\sum_{i} n_i(x)  \rho  n_{i}(x)^* }
{\sum_{i} \mathrm{trace}(\rho  n_{i}(x)^*  n_{i}(x) ) }
\end{equation}
and the distribution of the outcome is
\begin{equation}\label{e:instrpgen}
\pi(\mathrm d x;\rho,\mathcal N) 
    = \sum_{i} \mathrm{trace} (\rho  n_i(x)^*  n_i(x)) \nu (\mathrm d x).
\end{equation}
These formulae generalise naturally (\ref{e:instrp}) and (\ref{e:instrsigma}).
In the physics literature, this kind of representation is often called
the \emph{Kraus representation of a completely positive instrument}.
Space does not suffice to explain these terms,
in particular `complete positivity', further.
The interested reader is referred to \citet{davieslewis70}, 
\citet{davies76},  \citet{kraus83}, \citet{ozawa85}, 
\citet{nielsenchuang00}, \citet{loubenets01}, 
and \citet{holevo01book}. 

When the posterior state is disregarded, the instrument $\mathcal N$
gives rise to the measurement $M$ with density $m(x)=\sum_i n_i(x)^*  n_i(x)$
with respect to the dominating measure $\nu$. Clearly, a measurement $M$ can
be represented as the `data part' of an instrument in very many different ways.

Further results of \citet{ozawa85} generalise the realisability of 
measurements (Naimark, Holevo theorems) 
to the realisability of an arbitrary completely positive
instrument. Namely, after forming a compound system by taking the
tensor product with some ancilla, the instrument can be realised as a unitary 
(Schr\"odinger) evolution for some length of time,
followed by the action of a simple instrument 
(measurement of an observable, with state transition according to 
the L\"uders--von Neumann projection postulate). Therefore to say 
that the most general operation on a quantum system is a completely positive  instrument
comes down to saying: the only mechanisms
known in quantum mechanics are Schr\"odinger evolution, von Neumann 
measurement, and forming compound systems.
Combining these ingredients in arbitrary ways, one
remains within the class of completely positive instruments;
moreover, anything in that class can be realised in this 
way.

An instrument defined on one component of a product 
system can be extended in a natural way (similar to that described in 
Section \ref{ss:meas} for measurements) to an instrument on
the product system. Conversely, it is of great 
interest whether instruments on a product system can in some way be
reduced to `separate instruments on the separate sub-systems'. There 
are two important notions in this context. 
The first (similar to the concept 
of separability of measurements) is the \emph{mathematical}
concept of separability of 
an instrument defined on a product system: this is that each 
$n_i(x)$ in the Kraus representation of an instrument is a tensor product 
of separate matrices for each component. The second is the \emph{physical}
property which we shall call \emph{multi\-locality}: 
an instrument is called multi\-local, if it can be represented as a 
coarsening of a
composition of separate instruments applied sequentially to separate components of 
the product system, where the choice of each instrument at each stage 
may depend on the outcomes of the instruments applied previously. 
Moreover, each component of the system may be measured several times 
(i.e.\  at different stages), and the choice of component measured at the 
$n$th stage may depend on the outcomes at previous stages.
One should think of the different components of the quantum system as 
being localised at different locations in space. At each location 
separately, anything quantum is allowed, but all communication between 
locations is classical. It is a theorem of \citet{bennettetal99a} that 
every multi\-local instrument is separable, but that (surprisingly) not 
all separable instruments are multi\-local. It is an open problem to 
find a physically meaningful characterisation of separability, and 
conversely to find a mathematically convenient characterisation of 
multi\-locality. (Note our terminology is not standard: the word 
`unentangled' is used by some authors instead of `separable', and 
`separable' instead of `multi\-local').

Not all joint measurements (by which we just 
mean instruments on product systems), are separable, let alone 
multi\-local. Just as quantum randomised measurements can give strictly 
more powerful ways to probe the state of a quantum system than 
(combinations of) simple measurements and classical 
randomisation, so non-separable measurements can do strictly
better than separable
measurements at extracting information from product systems, even if 
a priori there is no interaction of any kind between the 
subsystems.

\section{Parametric Quantum Models and Likelihood}\label{s:lik}

A measurement from a quantum statistical model $(\boldrho  ,m)$ 
results in an observation $x$ with density
$$
p(x;\theta)=\mathrm{trace}(\rho(\theta)m(x))
$$
and log likelihood
$$
l(\theta)=\log\mathrm{trace}(\rho(\theta)m(x))
\thinspace .
$$

For simplicity, let us suppose that $\theta$ is one-dimensional.
It is often useful to express the log likelihood derivative in terms of the 
\emph{symmetric logarithmic derivative}
or \emph{quantum score} of $\boldrho$, 
denoted by $\rho_{/\!\!/\theta}$. This is defined implicitly as 
the self-adjoint solution of the equation
\begin{equation}
\rho_{/\theta}=\rho\circ\rho_{/\!\!/\theta}
\thinspace , 
\label{SLD}
\end{equation}
where $\circ$ denotes the Jordan product, i.e.
$$
\rho\circ\rho_{/\!\!/\theta}=
{\textstyle{\frac12}}(\rho\rho_{/\!\!/\theta}+\rho_{/\!\!/\theta}\rho)
\thinspace ,
$$
$\rho_{/\theta}$ denoting the ordinary derivative of $\rho$ with respect to
$\theta$ (term-by-term differentiation in matrix representations of $\rho$).
(We shall often suppress the argument $\theta$ in quantities like 
$\rho$, $\rho_{/\theta}$, $\rho_{/\!\!/\theta}$, etc.) 
The quantum score
exists and is essentially unique subject only to mild conditions. 
(For a discussion of this see, for example, p.\ 274 of \citealt{holevo82}.)

The likelihood score $l_{/\theta}(\theta)=(\mathrm d/\mathrm d\theta)l(\theta)$ 
may be expressed in terms of the
quantum score $\rho_{/\!\!/\theta}(\theta)$ of 
$\rho(\theta)$ as
\begin{align*}
l_{/\theta} (\theta ) ~&=~ p(x;\theta)^{-1}
\mathrm{trace}(\rho_{/\theta} (\theta ) m(x))  \\
~&=~ p(x;\theta)^{-1}{\textstyle{\frac12}}\mathrm{trace} \, 
((\rho (\theta ) \rho_{/\!\!/\theta} (\theta )
+ \rho_{/\!\!/\theta} (\theta ) \rho (\theta )) m(x)) 
 \\
~&=~ p(x;\theta)^{-1}\Re  \, \mathrm{trace} \, 
(\rho (\theta ) \rho_{/\!\!/\theta} (\theta ) m(x)) 
\thinspace , 
\end{align*}
where we have used the fact that for any self-adjoint matrices 
$P,Q,R$ and any matrix $T$ 
the trace operation satisfies 
$\mathrm{trace}(PQR)=\overline{\mathrm{trace}
(RQP)}$ and 
$\Re \, \mathrm{trace}(T)=
\frac{1}{2} \mathrm{trace}(T+T^{\ast})$.
It follows that
$$
\mathrm E_{\theta}(l_{/ \theta}(\theta )) = 
\mathrm{trace} ( \rho (\theta ) \rho_{/\!\!/\theta} (\theta ))
\thinspace .
$$
Thus, since the mean value of $l_{/ \theta}$ is $0$, we find that
\begin{equation}
\mathrm{trace}( \rho (\theta ) \rho_{/\!\!/\theta} (\theta )) =0
\thinspace .\label{3.3}
\end{equation}
The \emph{expected} (\emph{Fisher}) information
$i(\theta)=i(\theta;M)=\mathrm E_{\theta}(l_{/ \theta}(\theta )^{2})$ may be written as
\begin{equation}
i(\theta;M)=\int p(x;\theta)^{-1}\left(  
\Re \, \mathrm{trace}(\rho (\theta ) 
\rho_{/\!\!/\theta} (\theta ) m(x))\right)  ^{2}\nu(\mathrm d x)
\thinspace . \label{expinfo}
\end{equation}
It plays a key role in the quantum context, just as in classical
statistics, and is discussed in Section \ref{s:info}. In particular,
we shall discuss there its relation with the expected or Fisher
\emph{quantum information}
\begin{equation}\label{Qinfo}
I(\theta)=\mathrm{trace} ( \rho (\theta ) 
\rho_{/\!\!/\theta}(\theta)^{2} ).
\end{equation}
The quantum score is a self-adjoint operator, and therefore may be
interpreted as an observable which one might measure on the quantum 
system.
What we have just seen is that the outcome of a measurement of
the quantum score has mean zero and variance equal to the 
quantum Fisher information.

\section{Quantum Exponential and Quantum Transformation Models}
\label{s:qeqtm}

In traditional statistics, the two major classes of parametric models are the
exponential models (in which the log densities are affine functions of
appropriate parameters) and the transformation (or group) models (in which a
group acts in a consistent fashion on both the sample space and 
the parameter space); see \citet{barndorffnielsencox94}. The intersection
of these classes is the class of exponential transformation models, and its
members have a particularly nice structure. There are quantum analogues of
these classes, and they have useful properties. Below we outline some of these
briefly. Considerably more discussion is given in our works \citet{barndorffnielsenetal01,
barndorffnielsenetal01b,barndorffnielsenetal03}.

\subsection{Quantum Exponential Models}\label{ss:qem}

A \emph{quantum exponential model} is a quantum statistical model for which
the states $\rho(\theta)$ can be represented in the form
$$
\rho(\theta)= e^{-\kappa(\theta)} 
e^{\frac12\overline{\gamma}^{r}(\theta) 
T_{r}^{*}}\rho
_{0}e^{\frac12\gamma^{r}(\theta) T_{r}} \qquad \theta \in \Theta 
\thinspace , 
$$
where $\gamma=(\gamma^{1},\ldots,\gamma^{k}): \Theta\rightarrow\mathbb C ^{k}$, 
$T_{1}, \dots,T_{k}$ are
$d\times d$ matrices, $\rho_{0}$ is self-adjoint and non-negative
(but not necessarily a density matrix), the Einstein
summation convention (of summing over any index which appears as both a
subscript and a superscript) has been used, and $\kappa (\theta)$ is a 
log norming constant, given by
$$
\kappa (\theta) = \log \mathrm{trace} ( 
e^{\frac12\overline{\gamma}^{r}(\theta) 
T_{r}^{*}
}\rho_{0} e^{\frac12\gamma^{r}(\theta)T_{r}}) 
\thinspace .
$$

Three important special types of quantum exponential 
model are those in which $T_{1},\dots,T_{k}$ are self-adjoint 
(and for the first type, $T_{0}$, $T_{1}$, \dots,$T_{k}$ all commute),
and the quantum states have the forms
\begin{align}
\rho({\theta}) ~&=~ e^{-\kappa(\theta)}
\exp\left( T_0 + \theta^{r}T_{r} \right) ,
\label{mech} \\
\rho({\theta}) ~&=~ e^{-\kappa(\theta)}
\exp\left( {\textstyle{\frac12}}\theta^{r}T_{r}\right) \rho_{0} 
\exp\left( {\textstyle{\frac12}}\theta^{r}T_{r}\right) , 
\label{opsymm} \\
\rho({\theta}) ~&=~ \exp\left( -i{\textstyle{\frac12}}\theta^{r}T_{r}\right)  
\rho_{0}
\exp\left(  i{\textstyle{\frac12}}\theta^{r}T_{r}\right) 
\thinspace , \label{unitary} 
\end{align}
respectively, where $\theta=(\theta^{1},\dots,\theta^{k}) \in 
\mathbb R ^{k}$ and $\rho_{0}$ is nonnegative (and self-adjoint),
and the summation convention is in force.

We call these three types the quantum exponential models of 
\emph{mechanical} type, \emph{symmetric} type, and \emph{unitary} type 
respectively. The mechanical type arises (at least, with $k=1$) in
quantum statistical mechanics as a state of statistical equilibrium,
see \citet[{\ }Sect.\ 2.4.2]{gardinerzoller00}.
The symmetric type has theoretical statistical 
significance, as we shall see, connected among other things
to the fact that the quantum 
score for this model is easy to compute explicitly. The unitary type 
has physical significance connected to the fact that it is also a 
transformation model (quantum transformation models are defined in 
the next subsection). The mechanical type is a special case of the
symmetric type when $T_{0}$, $T_{1}$, \dots,$T_{k}$ all commute.

In general, the statistical model obtained by applying a 
measurement to a quantum exponential model is not an 
exponential model (in the classical sense). However, 
for a quantum exponential model of the form (\ref{opsymm}) in which 
\begin{equation}
T_{j} = t_{j}(X) \quad j = 1, \dots, k \qquad 
\mbox{for some self-adjoint $X$}
\thinspace ,
\label{Tcommute}
\end{equation}
i.e.\  the $T_{j}$ commute, the statistical model 
obtained by applying the measurement $X$
is a full exponential model. Various pleasant properties of such 
quantum exponent\-ial models then follow from standard 
properties of the full exponent\-ial models. 

The classical Cram\'{e}r--Rao bound for the variance of an 
unbiased estimator $t$ of $\theta$ is 
\begin{equation}
\mathrm{Var}(t)\geq i(\theta;M)^{-1}
\thinspace . \label{classCRB}
\end{equation}
Combining (\ref{classCRB}) with \citeauthor{braunsteincaves94}'
(\citeyear{braunsteincaves94})
quantum information bound $i(\theta;M)\le I(\theta)$
(see (\ref{qinfoineq}) in Section \ref{ss:infoclass}) 
yields  \citeauthor{helstrom76}'s (\citeyear{helstrom76}) quantum 
Cram\'{e}r--Rao bound 
\begin{equation}
\mathrm{Var}(t)\geq I(\theta)^{-1}
\thinspace,  \label{QCRB} 
\end{equation}
whenever $t$ is an unbiased estimator based on a quantum measurement.
It is a classical result that,
under certain regularity conditions, the following are equivalent:
(i) equality holds in (\ref{classCRB}), (ii) the score is
an affine function of $t$, (iii) the model is exponential
with $t$ as canonical statistic 
(cf.\ pp.\ 254--255 of \citealt{coxhinkley74}).
This result has a quantum analogue, see Theorems \ref{t:new2} 
and \ref{t:peter} and Corollary \ref{c:peter} below,
which states that under certain regularity conditions, 
there is equivalence between (i) equality holds in 
(\ref{QCRB}) for some unbiased estimator $t$ based on some 
measurement $M$,
(ii) the symmetric quantum score is an affine function of
commuting $T_{1},\dots,T_{k}$, and (iii) the quantum model is a
quantum exponential model of type (\ref{opsymm}) where
$T_{1},\dots,T_{k}$ satisfy (\ref{Tcommute}).
The regularity 
conditions which we assume below are indubitably too strong: the 
result should be true under minimal smoothness assumptions.

\subsection{Quantum Transformation Models}\label{ss:qtm}

Consider a parametric quantum model $(\boldrho , M)$ 
consisting of a parametrised family
$\boldrho =( \rho (\theta ) : \theta \in \Theta )$ of states and a 
measurement $M$ with outcome space 
$(\mathcal{X},\mathcal{A})$. Suppose that there exists a group, $G$, with 
elements $g$, acting both on $\mathcal X$ and on $\Theta$ in such a 
way that the following consistency condition (`equivariance') holds
\begin{equation}
\mathrm{trace} (\rho(\theta) M(A)) = \mathrm{trace} ( \rho(g\theta) M(gA)) 
\label{QTM}
\end{equation}
for all $\theta$, $A$ and $g$. If, moreover, $G$ acts transitively 
on $\Theta$ then we say that $(\boldrho, M)$ is a \emph{quantum transformation 
model}. In this case, the resulting statistical model for the outcome 
of a measurement of $M$, i.e. $(\mathcal X,\mathcal A, \mathcal P)$, 
where $\mathcal P=(\mathrm{trace}(\rho(\theta)M):\theta\in\Theta)$, 
is a classical transformation model. 
Consequently, the Main Theorem for transform\-ation
models, see \citet[{\ }pp.\ 56--57]{barndorffnielsencox94} and references given
there, applies to $(\mathcal{X}, \mathcal{A}, \mathcal{P})$.
Of special physical interest is the case in which 
the group acts on the states as a group of unitary
matrices. 

\begin{example}[Spin-half: great circle model]\label{e:greatcircle}
Consider 
the spin-half model $\rho(\theta)=U \,
\frac12(\boldone+\cos\theta \sigma_{x}+\sin\theta\sigma_{y})\,U^{*}$ 
where $U$ is a fixed $2\times 2$ unitary matrix, and $\sigma_{x}$ 
and $\sigma_{y}$ are two of the Pauli spin matrices, while the 
parameter $\theta$ varies through $[0,2\pi)$; see Subsection 
\ref{ss:spin12}. The matrix $U$ can 
always be written as $\exp(-i\phi\vec u\cdot\vec\sigma)$ for some real 
three-dimensional unit vector $\vec u$ and angle $\phi$. Considered 
as a curve on the Poincar\'e sphere, the model forms a great circle. 
If $U$ is the identity (or, equivalently, $\phi=0$) the curve just follows 
the equator; the presence of $U$ corresponds to rotating the sphere 
carrying this curve about the direction $\vec u$ through an angle $\phi$.
Thus our model describes an arbitrary great circle on the Poincar\'e 
sphere, parameterised in a natural way. Since we can write 
$\rho(\theta)=UV_{\theta}U^{*}\rho(0)UV_{\theta}^{*}U^{*}$, where the unitary 
matrix $V_{\theta}$ corresponds to rotation of the Poincar\'e sphere by an 
angle $\theta$ about the $z$-axis, we can write this model as a 
transformation model. 
The model is clearly also an exponential model of unitary type.
Perhaps surprisingly, it can be reparameterised so as also to be an 
exponential model of symmetric type. We leave the details (which 
depend on the algebraic properties of the Pauli spin matrices) to the 
reader, but just point out that a one-parameter pure-state exponential 
model of symmetric type has to be of the form 
$\rho(\theta)=\exp(-\kappa(\theta))\exp(\frac12 \theta\vec 
u\cdot\vec\sigma)\frac12(\boldone+\vec v\cdot\vec \sigma)
\exp(\frac12 \theta\vec 
u\cdot\vec\sigma)$ for some real unit vectors $\vec u$ and $\vec v$, 
since every self-adjoint $2\times2$ matrix is an affine function of
a spin matrix $\vec u\cdot \vec\sigma$. Now write out the 
exponential of a matrix as its power series, and use the fact that the 
square of any spin matrix is the identity. This example is due 
to \citet{fujiwaranagaoka95}.
\halmos\end{example}

\section{Quantum Exhaustivity, Sufficiency, and Quantum Cuts}\label{s:exhaust}

This section proposes and interrelates a number of concepts that will
constitute, we think, essential elements in the development of statistical
inference for the quantum context. The concepts are partly in the nature of
quantum analogues of key ideas of classical statistical inference, such as
sufficiency, the likelihood principle, etc. 

\subsection{Quantum Exhaustivity}\label{ss:exhaust}

Those quantum instruments for which no inform\-ation on the unknown 
parameter of a quantum parametric model of states can be obtained from 
subsequent measurements on the given physical system deserve special
attention. 
To simplify notation, we will write $\sigma(x;\theta , \mathcal{N})$
instead of $\sigma(x;\rho(\theta), \mathcal{N})$ for the posterior state.
We propose the following definition of \emph{exhaustivity}:

\begin{defn}[Exhaustive instrument]
A quantum instrument $\mathcal{N}$ 
is \emph{exhaustive} for a para\-meterised set 
$\boldrho = ( \rho (\theta) : \theta \in  \Theta )$ of states if
for all $\theta$ in $\Theta$ and for $\pi(\cdot ;\theta)$-almost all $x$, 
the posterior state $\sigma(x;\theta , \mathcal{N})$ does not depend on $\theta$. 
\halmos\end{defn}

Thus the posterior states obtained from exhaustive quantum instruments are
completely determined by the result of the measurement and do not depend on 
$\theta$. 

A useful strong form of exhaustivity is defined as follows.

\begin{defn}[Completely exhaustive instrument]
A quantum instrument $\mathcal{N}$ is \emph{completely exhaustive} if it is 
exhaustive for all parameterised sets of states. 
\halmos\end{defn}

Recall that any completely positive instrument---in other words, 
any physically realisable instrument---has posterior states given by (\ref{e:instrsigmagen})
and outcome distributed as (\ref{e:instrpgen}).
The following Proposition (which is a slight generalisation of a result of 
\citealt{wiseman99}) shows one way of constructing completely exhaustive 
completely positive quantum instruments.

\begin{prop}
Let the quantum instrument $\mathcal{N}$ be as above, with
$n_i(x)$ of the form
\begin{equation}
n_i(x) = \ketbra{\phi_{i,x}}{\psi_{x}}    
\thinspace ,
\label{Wcomplete}
\end{equation}
for some functions $(i,x) \mapsto\phi_{i,x}$ and $x \mapsto\psi_x$.
Then $\mathcal{N}$ is completely exhaustive. 
\end{prop}

\begin{proof}
By inspection we find that the posterior state is
$$
\sigma(x;\rho ,\mathcal{N}) = \frac {\sum_{i}  \ketbra{ \phi_{i,x} } { \phi_{i,x} } }
{\sum_{i} \braket{\phi_{i,x}}{ \phi_{i,x} }},
$$
which does not depend on the prior state $\rho$.
\myendproof\end{proof}

\subsection{Quantum Sufficiency}\label{ss:suff}

Suppose that the measurement $M^{\prime }=M\circ T^{-1}$ is a coarsening of
the measurement $M$. In this situation we say that 
$M^{\prime }$\ is \emph{(classically) sufficient} for $M$ with respect to a 
family of
states $\boldrho =(\rho (\theta ):\theta \in \Theta )$ on $\mathcal{H}$ 
if the mapping $T$ is sufficient for the identity mapping on $(\mathcal{X},
\mathcal{A})$ with respect to the family $(P(\cdot ;\theta ;M):\theta \in
\Theta )$ of probability measures on $(\mathcal{X},\mathcal{A})$ induced by 
$M$ and $\boldrho$.

As a further step towards a definition of quantum sufficiency, we introduce a
concept of inferential equivalence of parametric models of states.

\begin{defn}[Inferential equivalence]
Two parametric families of
states $\boldrho  =(\rho (\theta ):\theta \in \Theta )$ and 
$\boldsigma =(\sigma (\theta ):\theta \in \Theta )$ on Hilbert spaces 
$\mathcal{H}$ and $\mathcal{K}$\ are said to be 
\emph{inferentially equivalent} if for every measurement $M$ on $\mathcal{H}$
there exists a measurement $M'$ on $\mathcal{K}$\ such that for all $\theta
\in \Theta $ 
\begin{equation}\label{infequiv}
\mathrm{trace}(M(\cdot )\rho (\theta ))
= \mathrm{trace}(M'(\cdot )\sigma (\theta ))
\end{equation}
and vice versa.
(Note that, implicitly, the outcome spaces of $M$ and $M'$ are assumed to be
identical.)
\halmos\end{defn}

In other words, $\boldrho$ and $\boldsigma$ are equivalent if
and only if they give rise to the same class of possible classical models
for inference on the unknown parameter.

\begin{rmk} It is of interest to find characterisations of
inferential equivalence. This is a nontrivial problem, even when 
the Hilbert spaces $\mathcal{H}$\ and $\mathcal{K}$ are the same.
\halmos\end{rmk}

Next, let $\mathcal{N}$ denote a quantum instrument on a Hilbert space $\mathcal{H}$
and with outcome space $(\mathcal{X},\mathcal{A})$ and let $\mathcal{N}
^{\prime }=\mathcal N\circ T^{-1}$ be a coarsening
of $\mathcal{N}$ with outcome space 
$(\mathcal{Y},\mathcal{B})$, generated by a mapping $T$ from $(\mathcal{X},
\mathcal{A})$ to $(\mathcal{Y},\mathcal{B})$.
It is easy to show that  
the posterior states for the two instruments are related by
$$
\sigma (t;\theta, \mathcal{N}' )=\int_{T^{-1}(t)}\sigma(x;\theta, \mathcal{N} )
\pi(\mathrm{d }x|t;\theta, \mathcal{N}) ,
$$
where $\pi(\mathrm{d }x|t;\theta, \mathcal{N})$ is the conditional distribution 
of $x$ given $T(x)=t$ computed from 
$\pi(\mathrm{d }x;\theta, \mathcal{N})$.

\begin{defn}[Quantum sufficiency of instruments] Let 
$\mathcal{N}^{\prime }$ be a coarsening of an instrument $\mathcal{N}$
by $T : (\mathcal{X}, \mathcal{A}) 
\rightarrow (\mathcal{Y}, \mathcal{B})$. Then 
$\mathcal{N}^{\prime }$ is said to be \emph{quantum sufficient}\ with respect
to a family of states $(\rho (\theta ):\theta \in \Theta )$\ if
\begin{enumerate}
\item[(i)]  the measurement $M^{\prime}$ determined by 
$\mathrm{trace} (M^{\prime}(\cdot ) \rho) =\pi ( \cdot ; \rho, \mathcal{N}^{\prime})$
is sufficient for the measurement $M$ determined by
$\mathrm{trace} (M(\cdot ) \rho) =\pi ( \cdot ; \rho, \mathcal{N})$, 
with respect to the family $(\rho (\theta ):\theta \in \Theta )$,
\item[(ii)] for any $x\in \mathcal{X}$, 
the posterior families $(\sigma(x; \theta, \mathcal{N}):\theta \in \Theta)$ 
and $(\sigma(T(x);\theta, \mathcal{N}' ):\theta \in \Theta )$ are inferentially
equivalent.
\end{enumerate}
\halmos\end{defn}

\subsection{Quantum Cuts and Likelihood Equivalence}

In the theory of classical statistical inference, many important
concepts (such as sufficiency, ancillarity and cuts) can be expressed
in terms of the decomposition by a measurable function
$T : (\mathcal{X}, \mathcal{A}) \rightarrow (\mathcal{Y},
\mathcal{B})$
of each probability
distribution on $(\mathcal{X}, \mathcal{A})$ into the corresponding
marginal
distribution of $T(x)$
and the family of conditional distributions of $x$ given $T(x)$.
We now define analogous concepts in quantum statistics based on the
decomposition
\begin{equation}
\rho \mapsto (\pi ( \cdot ; \rho, \mathcal{N}), \sigma (\cdot ; \rho
, \mathcal{N}))
\label{qdecomp}
\end{equation}
by a quantum instrument $\mathcal{N}$ of each state $\rho$
into a probability distribution on $(\mathcal{X}, \mathcal{A})$ and a family of posterior states; 
see Section \ref{ss:instru}.

The classical concept of a cut encompasses those of sufficiency and
ancillarity and is therefore more basic.
A measurable function $T$ is a
{\it cut} for a set $\mathcal{P}$ of probability distributions on
$\mathcal{X}$ if for all $p_1$ and $p_2$ in
$\mathcal{P}$, the distribution on $\mathcal{X}$ obtained by
combining the marginal distribution of $T(x)$ given by $p_1$
with the family of conditional distributions of $x$ given $T(x)$
given by $p_2$ is also in $\mathcal{P}$; 
see, e.g.\ \citet[{\ }p.\ 38]{barndorffnielsencox94}.
Recent results on cuts for exponential models can be found in
\citet{barndorffnielsenkoudou95}, which also gives references
to the useful role which cuts have played in graphical models.
A generalisation to {\it local cuts} has become important
in econometrics \citep{christensenkiefer94, christensenkiefer00}.
Replacing the decomposition into marginal and conditional
distributions in the definition of a cut by the decomposition
(\ref{qdecomp}) yields the following quantum analogue.

\begin{defn}[Quantum cut]
A quantum instrument $\mathcal{N}$ is said to be a \emph{quantum cut}
for a family $\boldrho$ of states if for all
$\rho_1$ and $\rho_2$ in $\boldrho$
\begin{align*}
\pi ( \cdot ; \rho_3, \mathcal{N}) ~&=~
\pi ( \cdot ; \rho_1, \mathcal{N})  \\
\sigma (\cdot ; \rho_3 , \mathcal{N}) ~&=~
\sigma (\cdot ; \rho_2 , \mathcal{N}) . 
\end{align*}
for some $\rho_3$ in $\boldrho$.
\halmos\end{defn}
   
Thus, if $\mathcal{N}$ is a quantum cut for a family
$\boldrho = ( \rho( \theta ) : \theta  \in \Theta )$
with $\rho$ a one-to-one function then
$\Theta$ has the product form $\Theta = \Psi \times \Phi$ and furthermore
$\sigma (\cdot ; \rho( \theta ), \mathcal{N})$ depends on $\theta$
only through $\psi$, and $\pi (\cdot ; \rho( \theta ),\mathcal{N})$
depends on $\theta$ only through $\phi$.

Since a quantum instrument $\mathcal{N}$ is exhaustive for a parameterised set
$\boldrho = ( \rho( \theta ) : \theta  \in \Theta )$ of states
if the family $\sigma (\cdot ; \rho (\theta ) , \mathcal{N})$ of
posterior states does not depend on $\theta$, exhaustive quantum
instruments are quantum cuts of a special kind. They can be regarded
as quantum analogues of sufficient statistics.
At the other extreme are the quantum instruments for which the
distributions $\pi (\cdot ; \rho (\theta ) , \mathcal{N})$ do not depend
on $\theta$. These can be regarded as quantum analogues of ancillary
statistics.

Unlike exhaustivity, the concept of quantum sufficiency involves not
only a quantum instrument but also a coarsening. The definition of
quantum sufficiency can be extended to the following version
involving parameters of interest.

\begin{defn}[Quantum sufficiency for interest parameters]
Let $\boldrho = ( \rho (\theta ) : \theta \in  \Theta )$ be a
family of states and let $\psi : \Theta \rightarrow \Psi$ map $\Theta$
to the space $\Psi$ of interest parameters.
A coarsening $\mathcal{N}'$ of a quantum instrument $\mathcal{N}$ by a
mapping $T$ is said to be \emph{quantum sufficient} for $\psi$ on
$\boldrho$ if
\begin{enumerate}
\item[(i)] the measurement $M^{\prime}$ determined by 
$\mathrm{trace} (M^{\prime}(\cdot ) \rho) =\pi ( \cdot ; \rho, \mathcal{N}^{\prime})$
is sufficient for the measurement $M$ determined by
$\mathrm{trace} (M(\cdot ) \rho) =\pi ( \cdot ; \rho, \mathcal{N})$, 
with respect to the family $\boldrho$,   
\item[(ii)] for all $\theta_1$ and $\theta_2$ with $\psi (\theta_1)
= \psi (\theta_2)$ and for all $x$ in $\mathcal{X}$, the sets
$\sigma (x; \rho (\theta _1), \mathcal{N})$ and
$\sigma (T(x); \rho (\theta _2), \mathcal{N}')$ of posterior states
are
inferentially equivalent.
\end{enumerate}
\halmos\end{defn}
   
Consideration of the likelihood function obtained by applying a
measurement to a parameterised set of states suggests that the
following weakening of the concept of inferential equivalence may be useful.

\begin{defn}[Strong likelihood equivalence]
Two parametric families of states
$\boldrho = ( \rho (\theta ) : \theta \in \Theta )$ and 
$\boldsigma =( \sigma (\theta ) : \theta \in \Theta )$ on
Hilbert spaces
$\mathcal{H}$ and $\mathcal{K}$ respectively are said to be
\emph{strongly likelihood equivalent} if for every measurement $M$ on
$\mathcal{H}$ 
there is a measurement $M'$ on $\mathcal{K}$ with the same outcome 
space, 
such that
$$
\frac{ {\rm trace} ( M(\mathrm{d} x)  \rho (\theta )) }
{ {\rm trace} ( M(\mathrm{d} x)  \rho (\theta ')) }
= \frac{ {\rm trace} ( M'(\mathrm{d} x) \sigma (\theta )) }
{ {\rm trace} ( M'(\mathrm{d} x) \sigma (\theta ')) }
\qquad  \theta , \theta ' \in \Theta
$$
(whenever these ratios are defined) and vice versa.
\halmos\end{defn}

Thus the likelihood function of the statistical model obtained by
applying $M$ to $\boldrho$ is equivalent to that obtained by
applying $M'$ to $\boldsigma$, 
for the same outcome of each instrument.

Consideration of the distribution of the likelihood ratio leads to 
the following definition.

\begin{defn}[Weak likelihood equivalence]
Two parametric families of states
$\boldrho = ( \rho (\theta ) : \theta \in \Theta )$ and 
$\boldsigma = ( \sigma (\theta ) : \theta \in \Theta)$ on
Hilbert spaces
$\mathcal{H}$ and $\mathcal{K}$ respectively are said to be
\emph{weakly likelihood equivalent} if for every measurement $M$ on
$\mathcal{H}$ with outcome space $\mathcal{X}$ 
there is a measurement $M'$ on $\mathcal{K}$ with some outcome space 
$\mathcal{Y}$ such that the likelihood ratios 
$$
\frac{{\rm trace} ( M(\mathrm{d} x)  \rho (\theta ))}
{{\rm trace} ( M(\mathrm{d} x)  \rho (\theta '))}
\qquad  \rm{and} \qquad  \frac{{\rm trace} ( M'(\mathrm{d} y) \sigma (\theta ))}
{{\rm trace} ( M'(\mathrm{d} y) \sigma (\theta '))}
$$
have the same distribution for all $\theta , \theta '$ in $\Theta$,
and vice versa.
\halmos\end{defn}

The precise connection between likelihood equi\-valence and
inferential equivalence is not yet known but the following
conjecture appears reasonable.

\vskip 12pt
\noindent
{\bf Conjecture.}
Two quantum models are strongly likelihood equivalent if and only if they are
inferentially equivalent
up to quantum randomisation.

\section{Quantum and Classical Fisher Information}\label{s:info}

In Section \ref{s:lik} we showed how to express the Fisher information 
in the outcome of a measurement in terms of the quantum score. 
In this section we discuss quantum analogues of Fisher 
information and their relation to the classical concepts.

\subsection{Definition and First Properties}\label{ss:infodef}

Differentiating (\ref{3.3}) with respect to $\theta$, writing $\rho
_{/\!\!/\theta/\theta}$ for the derivative of the symmetric 
logarithmic derivative
$\rho_{/\!\!/\theta}$ of $\rho$, and using the defining 
equation (\ref{SLD}) for $\rho_{/\!\!/\theta}$, we obtain
\begin{align*}
0 ~&=~ \mathrm{trace}
(\rho_{/\theta}(\theta) \rho_{/\!\!/\theta}(\theta)
+ \rho (\theta) \rho_{/\!\!/\theta/\theta}(\theta))  \\
~&=~ \mathrm{trace}\left(  
{\textstyle{\frac12}}\Bigl(\rho (\theta) \rho_{/\!\!/\theta}(\theta)
+\rho_{/\!\!/\theta}(\theta) \rho(\theta)\Bigr) \rho_{/\!\!/\theta}(\theta)
\right)  
+ \mathrm{trace}(\rho (\theta) \rho_{/\!\!/\theta/\theta}(\theta))  \\
~&=~ I(\theta)- \mathrm{trace} (\rho (\theta) J(\theta))
\thinspace , 
\end{align*}
where
$$
I(\theta)=\mathrm{trace} \left( \rho (\theta ) \rho_{/\!\!/\theta}(\theta)^{2} \right) 
$$
is the \emph{expected} (or \emph{Fisher}) \emph{quantum information},
already mentioned in Sections \ref{s:lik} and \ref{s:qeqtm}, and
$$
J(\theta)=-\rho_{/\!\!/\theta/\theta} (\theta)
\thinspace , 
$$
which we shall call the \emph{observable quantum information}. 
Thus
$$
I(\theta)=\mathrm{trace} \left( \rho (\theta) J(\theta) \right)
\label{I=EJ}
\thinspace ,
$$
which is a quantum analogue of the classical relation 
$i(\theta)=\mathrm E_{\theta} (j(\theta))$ between expected and observed 
information 
(where $j(\theta) = - l_{/ \theta / \theta}(\theta)$). 
Note that $J(\theta)$ is an observable, just as $j(\theta)$ is a random 
variable.

Neither $I(\theta)$ nor $J(\theta)$ depends on the choice of measurement,
whereas $i(\theta)=i(\theta;M)$ does depend on the measurement $M$. 
Expected quantum information behaves additively, i.e.\  
for parametric quantum models of states of the form
$\boldrho :\theta \mapsto 
\rho_{1}(\theta)\otimes\dots \otimes \rho_{n}(\theta)$
(which model `independent particles'), the associated 
expected quantum information satisfies
$$
I_{\rho_{1}\otimes\dots\otimes\rho_{n}}(\theta) = 
\sum_{i=1}^{n}I_{\rho_{i}}(\theta)
\thinspace , 
$$
which is analogous to the additivity property of Fisher information.

In the case of a multivariate parameter $\theta$, the 
\emph{expected quantum
inform\-ation} matrix $I(\theta)$ is defined in terms of the 
quantum scores by
\begin{equation}
I(\theta)_{jk} = {\textstyle{\frac12}} \mathrm{trace} \left( 
\rho_{/\!\!/\theta_{j}}(\theta) \rho (\theta) 
\rho_{/\!\!/\theta_{k}}(\theta) 
+ \rho_{/\!\!/\theta_{k}}(\theta) \rho (\theta)
\rho_{/\!\!/\theta_{j}}(\theta)
\right)
\thinspace .
\label{matrixI}
\end{equation}

\subsection{Relation to Classical Expected Information}
\label{ss:infoclass}

Suppose that $\theta$ is one-dimensional.
There is an important relationship between expected quantum
inform\-ation $I(\theta)$ and classical expected information
$i(\theta;M)$, due to 
\citet{braunsteincaves94}, namely that for any measurement $M$ with density 
$m$ with respect to a $\sigma$-finite measure $\nu$ on $\mathcal{X}$,
\begin{equation}
i(\theta;M)\leq I(\theta)
\thinspace , \label{qinfoineq}
\end{equation}
with equality if and only if, for $\nu$-almost all $x$,
\begin{equation}
m(x)^{1/2}\rho_{/\!\!/\theta}(\theta) \rho (\theta)^{1/2} 
= r(x) m(x)^{1 /2} \rho (\theta) ^{1/2}
\thinspace ,
\label{QCREcond2}
\end{equation}
for some real number $r(x)$.

For each $\theta$, there are measurements which attain the bound in the
quantum information inequality (\ref{qinfoineq}). For instance, we can choose
$M$ such that each $m(x)$ is a projection onto an eigenspace of the quantum
score $\rho _{/\!\!/ \theta} (\theta)$. Note that this attaining measurement 
may depend on $\theta$.

\begin{example}[Information for spin-half]\label{e:spin12qi}
Consider a spin-half particle in the pure state
$\rho=\rho(\eta,\theta)=|\psi(\eta,\theta)\rangle\langle\psi(\eta,\theta)|$ 
given by
$$
|\psi(\eta,\theta)\rangle=\left(
\begin{matrix}
e^{-i\theta/2}\cos(\eta/2)\\
e^{i\theta/2}\sin(\eta/2)
\end{matrix}
\right)  
\thinspace .
$$
As we saw in Subsection \ref{ss:spin12}, $\rho$ can be written as 
$\rho=(\boldone +u_{x}\sigma
_{x}+u_{y}\sigma_{y}+u_{z}\sigma_{z})/2=
\frac{1}{2}(\boldone +\vec{u}\cdot\vec{\sigma})$, 
where $\vec{\sigma}=(\sigma_{x},\sigma_{y},\sigma_{z})$ are the 
three Pauli spin matrices and 
$\vec{u}=(u_{x},u_{y},u_{z})=\vec{u}(\eta,\theta)$ is
the point on the Poincar\'e sphere $S^{2}$ with polar coordinates
$(\eta,\theta)$. Suppose that the colatitude $\eta$ is known
and exclude the degenerate cases $\eta=0$ or $\eta=\pi$; the 
longitude $\theta$ is the unknown parameter.

Since all the $\rho(\theta)$ are pure, one can show that
$\rho_{/\!\!/\theta}(\theta) =2\rho_{/\theta}(\theta)
=\vec{u}_{/\theta}(\theta)\cdot\vec{\sigma} =
\sin(\eta)\,\vec{u}(\pi/2,\theta+\pi/2)\cdot \vec{\sigma}$.
Using the properties of the Pauli matrices, one finds that 
the quantum information is 
\[
I(\theta)=\mathrm{trace}(\rho (\theta) \rho_{/\!\!/\theta}(\theta)^{2}) =
\sin^{2}\eta.
\]
Summarising some results from \citet{barndorffnielsengill00}, 
we now discuss a condition that a measurement must satisfy in order for 
it to achieve this information.

It follows from (\ref{QCREcond2}) that, for a pure 
spin-half state $\rho=|\psi\rangle\langle \psi|$, a necessary and sufficient
condition for a measure\-ment to achieve the information bound is: 
for $\nu$-almost all $x$, $m(x)$ is proportional to
a one-dimensional projector $|\xi(x)\rangle\langle\xi(x)| $ satisfying
$$
\langle\xi(x)|2\rangle\langle2|a\rangle ~ = ~ r(x)\langle\xi(x)|1\rangle
\thinspace ,
$$
where $r(x)$ is real, $|1\rangle=|\psi\rangle$, $|2\rangle=|\psi\rangle ^{\perp}$ 
($|\psi \rangle ^{\perp}$ being a unit vector in $\mathbb C ^2$
orthogonal to $|\psi \rangle$)
and $|a\rangle=2|{\psi}\rangle_{/ \theta}$. It can be seen that geometrically 
this means that $|\xi(x)\rangle$ corresponds to a point 
on $S^{2}$ in the plane
spanned by $\vec{u}(\theta)$ and $\vec{u}_{/\theta}(\theta)$.

If $\eta\ne\pi/2$, then distinct values of $\theta$ give distinct
planes, and all these planes intersect in the origin only. Thus no
single measurement $M$ can satisfy $I(\theta ) = i(\theta ; M)$
for all $\theta$. On the other hand, if
$\eta=\pi/2$, so that the states $\rho(\theta)$ lie on a
great circle in the Poincar\'{e} sphere, then the planes defined for
each $\theta$ are all the same. In this case {\em any}
measurement $M$ with all components proportional to projector 
matrices for directions in the plane $\eta=\pi/2$ satisfies 
$I(\theta ) = i(\theta ; M)$ for all $\theta \in \Theta$. In particular,
\emph{any simple measurement in that plane} has this property. 

More generally, a smooth one-parameter model of a spin-half pure 
state with every\-where positive quantum information admits a uniformly 
attaining measurement, i.e. such that $I(\theta ) = i(\theta ; M)$ for 
all $\theta \in \Theta$, if and only if the model is a great circle on 
the Poincar\'e sphere. This is actually a quantum exponential 
transformation model, see Example \ref{e:greatcircle}.
\halmos\end{example}

When the state $\rho$ is strictly positive, and under further 
nondegeneracy conditions, essentially the only way to achieve the bound
(\ref{qinfoineq}) is through measuring the quantum score. 
In the discussion below we first keep the value of $\theta$ fixed.
Since any nonnegative self-adjoint matrix can be written as a 
sum of rank-one matrices (using its eigenvalue-eigenvector 
decomposition), it follows that any dominated measurement can be refined 
to a measurement for which each $m(x)$ is 
of rank $1$, thus $m(x)=r(x)|\xi(x)\rangle\langle\xi(x)| $
for some real $r(x)$ and state vector $|\xi(x)\rangle$,
see the end of Section \ref{ss:meas}.
If one measurement is the refinement of another, then the 
distributions of the outcomes are related in the same way. Therefore,
under refinement of a measurement, Fisher expected 
information cannot decrease. Therefore if any measurement achieves
(\ref{qinfoineq}), there is also a measurement with rank $1$ 
components achieving the bound. Consider such a measurement.
Suppose that $\rho$ is positive and that all the 
eigenvalues of $\rho_{/\!\!/\theta}$ are different. The
condition $m(x)^{1/2}\rho_{/\!\!/\theta}\rho^{1/2} 
=r(x) m(x)^{1 /2} \rho ^{1/2}$ is then 
equivalent to $|\xi(x)\rangle\langle \xi(x) |\rho_{/\!\!/\theta}
=r(x)|\xi(x)\rangle\langle \xi(x) |$, which states that
$\xi(x)$ is an eigenvector of $\rho_{/\!\!/\theta}$. Since we must 
have $\int m(x)\nu(\mathrm d x)=\boldone$, it follows that all 
eigenvectors of $\rho_{/\!\!/\theta}$ occur in this way in components $m(x)$ of 
$M$. The measurement can therefore be reduced or coarsened
to a simple measurement of the quantum score, and the reduction (at the 
level of the outcome) is sufficient.

Suppose now that the state $\rho(\theta)$ is strictly positive for all
$\theta$, and that the quantum score has distinct eigenvalues for at 
least one value of $\theta$.
Suppose that a single measurement exists attaining (\ref{qinfoineq}) 
uniformly in $\theta$. Any refinement of this measurement therefore 
also achieves the bound uniformly; in particular, the refinement to 
components which are all proportional to projectors onto orthogonal
one-dimensional eigenspaces of the quantum score at the value of 
$\theta$ where the eigenvalues are distinct does so. Therefore the 
eigenvectors of the quantum score at this value of $\theta$ are 
eigenvectors at all other values of $\theta$. Therefore there is a 
self-adjoint operator $X$ with distinct eigenvalues such that
$\rho_{/\!\!/\theta}(\theta)=f(X;\theta)$ for each $\theta$.
Fix $\theta_{0}$ and 
let $F(X;\theta)=\int_{\theta_{0}}^{\theta}f(X;\theta)\mathrm d\theta$.
Let $\rho_{0}=\rho(\theta_{0})$.
If we consider the defining equation (\ref{SLD}) as a differential 
equation for $\rho(\theta)$ given the quantum score, and with initial 
condition $\rho(\theta_{0})=\rho_{0}$, we see that a
solution is $\rho(\theta)=\exp(\frac12 
F(X;\theta))\rho_{0}\exp(\frac12 F(X;\theta))$.
Under smoothness conditions the solution is unique.
Rewriting the form of this solution, we come to the following theorem:

\begin{thm}[Uniform attainability of quantum information bound]\label{t:new}
Suppose that the state is every\-where positive, the quantum 
score has distinct eigenvalues for some value of $\theta$, and is 
smooth. Suppose that a measurement $M$ exists with $i(\theta;M)=I(\theta)$ 
for all $\theta$, thus attaining the Braunstein--Caves information bound 
(\ref{qinfoineq})
uniformly in $\theta$. Then there is an observable $X$ such that a simple 
measurement of $X$ also achieves the bound uniformly, and the model 
is of the form 
\begin{equation}
    \rho(\theta)=c(\theta)\exp({\textstyle \frac12} F(X;\theta))\rho_{0}\exp(
    {\textstyle \frac12}
    F(X;\theta))
\label{new}
\end{equation}
for a function $F$, indexed by $\theta$, of an observable $X$ where 
$c(\theta)=1/\mathrm{trace}(\rho_{0}\exp(F(X;\theta)))$, $\rho_{/\!\!/ 
\theta}(\theta)=f(X;\theta)-\mathrm{trace}(\rho(\theta) f(X;\theta))$, and
$f(X;\theta)=F_{/ \theta}(X;\theta)$.
Conversely, for a model of this form, a measurement of $X$ 
achieves the bound uniformly.
\end{thm}

\begin{rmk}[Spin-half case]
In the spin-half case, if the information is positive then the 
quantum score has distinct eigenvalues, since the outcome of a 
measurement of the quantum score always equals one of the eigenvalues,
has mean zero, and positive variance.
\halmos\end{rmk}

\begin{thm}[Uniform attainability of quantum Cram\'er--Rao bound]\label{t:new2}
Suppose that the positivity and nondegeneracy conditions of the previous 
theorem are satisfied, and suppose that for the outcome of some measurement 
$M$ there is a statistic $t$ such that,
for all $\theta$, $t$ is an unbiased estimator of $\theta$ achieving Helstrom's
quantum Cram\'er--Rao bound (\ref{QCRB}), $\mathrm{Var}(t) = I(\theta)^{-1}$.
Then the model is a quantum exponential model of symmetric 
type (\ref{opsymm}), 
\begin{equation*}
    \rho(\theta)=c(\theta)\exp({\textstyle{\frac12}} \theta T ) \rho_{0}\exp(
    {\textstyle{\frac12}}
    \theta T )
\end{equation*}
for some observable $T$, and simple measurement of $T$ is equivalent to
the coarsening of $M$ by $t$.
\end{thm}
\begin{proof}

The coarsening $M'=M\circ t^{-1}$ by $t$ of the measurement $M$ also 
achieves the quantum information bound (\ref{qinfoineq}) uniformly; i.e.\ 
$i(\theta;M')= I(\theta)$. Applying Theorem \ref{t:new} to this 
measurement, we discover that the model is of the form (\ref{new}),
while (if necessary refining the measurement to have rank one 
components) $t$ can be considered as a function of the outcome of a 
measurement of the observable $X$, and it achieves the classical 
Cram\'er--Rao bound for unbiased estimators of $\theta$ based on this outcome. 
The density of the outcome (with respect to counting measure on the 
eigenvalues of 
$X$) is found to be 
$c(\theta)\exp(F(x;\theta))\mathrm{trace}(\rho_{0}\Pi_{x})$
where $\Pi(x)$ is the projector onto the eigenspace of $X$ corresponding
to eigenvalue $x$.
Hence, up to addition of functions of $\theta$ or $x$ alone,
$F(x;\theta)$ is of the form $\theta t(x)$.
\myendproof\end{proof}

The basic inequality (\ref{qinfoineq}) holds also when the 
dimension of $\theta$ is greater than one. 
In that case, the quantum information matrix $I(\theta)$ is 
defined in (\ref{matrixI}) and the Fisher information matrix 
$i(\theta;M)$ is defined by
$$
i_{rs}(\theta;M) = \mathrm E_{\theta}(l_{r}(\theta)l_{s}(\theta))
\thinspace , 
$$
where $l_{r}$ denotes $l_{/\theta^{r}}$ etc. 
Then (\ref{qinfoineq}) holds in the sense that 
$I(\theta) -  i(\theta;M)$ is positive semi-definite. The inequality 
is sharp in the sense that $I(\theta)$ is the smallest matrix 
dominating all $i(\theta;M)$. However it is typically not attainable,
let alone uniformly attainable.

Theorem \ref{t:new} can be generalised to the case of a vector parameter. 
This also leads to a generalisation of Theorem \ref{t:new2}, which is the 
content of Corollary \ref{c:peter} below.

\begin{thm}\label{t:peter}
Let $(\rho(\theta):\theta\in\Theta)$ be a 
twice differentiable parametric quantum model.
If
\begin{enumerate}
\item[(i)]      there is a measurement $M$ with $i(\theta;M)=I(\theta)$ 
for all $\theta$, 
\item[(ii)]     $\rho (\theta )$ is positive for all $\theta$,
\item[(iii)]    $\Theta$ is simply connected
\end{enumerate}
then, for any $\theta _0$ in $\Theta$, there are an observable $X$ and a
function $F$ (possibly depending on $\theta _0$) such that
$$
\rho (\theta )  = \exp \left( {\textstyle \frac12} F(X; \theta) \right) 
\rho (\theta _0)
\exp \left( {\textstyle \frac12} F(X; \theta) \right) .
$$
\end{thm}

\begin{cor}\label{c:peter}If, under the conditions of Theorem 
\ref{t:peter}, there exists an unbiased estimator $t$ of $\theta$ 
based on the measurement $M$ achieving (\ref{QCRB}), then the model
is a quantum exponential family
of symmetric type (\ref{opsymm}) with commuting $T_{r}$.
\end{cor}

Versions of these results have been known for some time; 
see \citet{young75}, \citet{fujiwaranagaoka95}, \citet{amarinagaoka00};
compare especially our Corollary \ref{c:peter} to 
\citet[{\ }Theorem 7.6]{amarinagaoka00},
and our Theorem \ref{t:peter} to parts (I)--(IV) of the subsequent 
outlined proof in  \citet{amarinagaoka00}.
Unfortunately, precise regularity conditions and detailed proofs 
seem to be available elsewhere only in some earlier publications in Japanese.
Note that we have obtained the same conclusions, by a different proof, 
in the spin-half pure state case, Example \ref{e:spin12qi}. This 
indicates that a more general result is possible without the 
hypothesis of positivity of the state. See \citet{matsumoto02} for important
new work on the pure state case.

The symmetric logarithmic derivative is not the unique quantum 
analogue of the classical statistical concept of score.
Other analogues include the right, left and balanced 
derivatives obtained from suitable variants of (\ref{SLD}).
Each of these gives a quantum information inequality 
and a quantum Cram\'er--Rao bound analogous to 
(\ref{qinfoineq}) and (\ref{QCRB}). See \citet{belavkin76}, and (as yet) unpublished
new work by this author.
There is no general relationship between the 
various quantum information inequalities when the dimension of 
$\theta$ is greater than one.

Asymptotic optimality theory for quantum estimation has only just started to be developed;
see \citet{gillmassar00}, \citet{gill01b}, and \citet{keylwerner01}; for an application see
\citet{hannemannetal02}.

\section{Classical versus Quantum}\label{s:qu-class}

This section makes some general comments on the relation between
classical and quantum probability and statistics. This has been a matter of 
heated controversy ever since the discovery of quantum mechanics. 
It has mathematical, physical, and philosophical 
ingredients, and much confusion, if not controversy, has been generated by 
problems of interdisciplinary communication between mathematicians, 
physicists, philosophers and more recently statisticians. 
Authorities from both physics and mathematics, 
perhaps starting with \citet{feynman51},
have promoted vigorously the standpoint that `quantum probability' 
is something very different from `classical probability'.
\citet{malleyhornstein93} conclude from a perceived conflict between 
classical and quantum probability that 
`quantum statistics' should be set apart from classical statistics.
Even \citet{williams01} states that Nature chooses a different model
for probability for the quantum world than for the classical world.

In our opinion, though important mathematical and physical facts 
lie at the root of these statements, they are misleading, since they seem
to suggest that quantum probability and quantum statistics do not belong
to the field of classical probability and statistics. However, quantum probabilities
have the same meaning (whether you are a Bayesian or a frequentist) as classical
probabilities, and statistical inference problems from 
quantum mechanics fall squarely in the framework of classical statistics. 
The statistical design problems are special to the field.

Our stance is that the predictions which quantum mechanics makes of the 
real world are stochastic in nature. A quantum physical model of a 
particular phenomenon allows one to compute probabilities of all possible 
outcomes of all possible measurements of the quantum system. The word 
`probability' means here `relative frequency in many independent repetitions'.
The word `measurement' is meant in the broad sense of `macroscopic results of
interactions of the quantum system under study with the outside world'.
These predictions depend on a summary of the state of the quantum system.
The word `state' might suggest some fundamental property of a particular 
collection of particles, but for our purposes all we need to understand 
under the word is `a convenient mathematical encapsulation of
the information needed to make any such predictions'. 
Some physicists argue that it is meaningless to talk
of the state of a particular particle, one can only talk of the state of a large 
collection of particles prepared in identical circumstances;
this is called a \emph{statistical ensemble}. Others take the point of view 
that when one talks about the state of a 
particular quantum system one is really talking about a property of the 
mechanism which generated that system. 
Given that quantum mechanics predicts only probabilities, as far
as real-world predictions are concerned
the distinction between on the one hand a property of an 
ensemble of particles or of a procedure to prepare particles, and on 
the other hand a property of one particular particle,
is a matter of semantics.  However, if one would like to understand quantum 
mechanics by somehow finding a more classical (intuitive) physical theory in the 
background which would explain the observed phenomena, this becomes an 
important issue. It is also an issue for cosmology, when there is 
only one closed quantum system under study: the universe.
At this level there \emph{is} a remarkable difference between classical and
quantum probabilities:  according to the celebrated theorem of \citet{bell64},
it is impossible to derive the probabilities described in
quantum mechanics by an underlying deterministic theory from which the
probabilities arise `merely' as the reflection of statistical variation in the initial
conditions, unless one accepts grossly unphysical nonlocality in the
`hidden variables'. Thus quantum probabilities are fundamentally irreducible,
in contrast to every other physical manifestation of randomness known to us.

It follows from our standpoint that `quantum statistics' is classical 
statistical inference about unknown parameters in models for data arising 
from measurements on a quantum system. However, just as in biostatistics,
geostatistics, etc., many of these statistical problems have a common 
structure and it pays to study the core ideas and common features in detail. As we 
have seen, this leads to the introduction of mathematical objects such as 
quantum score, quantum expected information, quantum exponential family, 
quantum transformation model, quantum cuts, and so on; the names are deliberately chosen 
because of analogy and connections with the existing notions from classical 
statistics.

Already at the level of probability (i.e.\  before statistical considerations 
arise) one can see a deep and fruitful analogy between the mathematics of quantum states and 
observables on the one hand, and classical probability measures and random variables 
on the other. Note that collections of random variables 
and collections of operators can both be endowed with algebraic structure 
(sums, products, \dots). It is a fact that from an abstract point of view a basic structure 
in probability theory---a collection of random variables $X$ on a countably 
generated probability space, together with their expectations $\int X 
\mathrm d P$ under a given probability measure $P$---can be represented by a 
(commuting) subset of the set of self-adjoint operators $Q$ on a separable Hilbert space, 
together with the expectations $\mathrm{trace}(\rho Q)$ computed using the trace 
rule under a given state $\rho$. Thus a \emph{basic} structure in classical 
probability theory is isomorphic to a \emph{special case} of a basic structure in 
quantum probability. `Quantum probability', or `noncommutative 
probability theory' is the name of the branch of mathematics which takes as
its starting point  
the mathematical structure of states and observables in quantum mechanics.
From this mathematical point of view, one may claim that classical probability is a 
special case of quantum probability. The claim does entail, however, a rather narrow 
(functional analytic) view of classical probability. Moreover, many probabilists will feel that
abandoning commutativity is throwing away the baby with the 
bathwater, since this broader mathematical structure has no analogue of the
sample outcome $\omega$, and hence no opportunity for a
probabilist's beloved probabilistic arguments.

As statisticians, we would like to argue (tongue in cheek)
that quantum probability is merely a special case of classical statistics.  
A quantum probability model is determined by specifying the expectations of every observable.
This is equivalent to specifying a family of classical probability models: namely the joint
probability distribution of the measurements of every commuting subset of observables.
The basic structure of quantum probability is mathematically equivalent to a 
particular case of the basic structure of classical statistical inference---namely, 
an indexed family of probability models.

\section{Other Topics}\label{s:other}

There are many further topics in quantum physics where more extensive use of
knowledge and techniques from classical statistics and probability seems
likely to lead to substantial scientific advances. However, classical
concepts and results from the latter fields will often need considerable
modification or recasting to be suitable and relevant for the quantum world,
as is exemplified in parts of the previous Sections. 

Here we shall indicate briefly a few of the topics. The selection of these
is motivated mainly by our own current interests rather than by an aim to be in
some way representative of the broad picture. However, the topics listed are
all subject to considerable developments in the current literature. For more
detailed accounts, with references to the physics and mathematics
literature, see \citet*{barndorffnielsenetal01b,barndorffnielsenetal03}.

\subsection{Quantum Tomography}\label{ss:tomo}

In its simplest form, the problem of quantum tomography is as follows.

The simple harmonic oscillator is the basic model for the motion of a
quantum particle in a quadratic potential well on the real line. Precisely
the same mathematical structure describes oscillations of a single mode of
an electromagnetic field (a single frequency in one direction in space). In
this type of structure one considers the quadrature observable at phase 
$\phi $, given by $X_{\phi }=Q\cos \phi +P\sin \phi$, where $Q$ and $P$ are
the position and momentum operators. Here, the underlying Hilbert space 
$\mathcal H$ is infinite dimensional and the operators $Q$ and $P$ 
can be characterised abstractly by the commutation relation 
$[Q,P]=i\hbar\boldone$.

Given independent measurements of 
$X_{\phi }$, with $\phi $ drawn repeatedly at random from the uniform
distribution on $(0,2\pi ]$, the aim is to reconstruct the unknown state 
$\rho $ of the quantum system. In statistical terms, we wish to do
nonparametric estimation of $\rho $ from $n$ independent and identically
distributed observations $(\phi _{i},x_{i})$, with $\phi _{i}$ as just
described and $x_{i}$ from the measurement of $X_{\phi _{i}}$. In quantum
optics, measuring a single mode of an electromagnetic field in what is
called a quantum homodyne experiment, this would be the appropriate model
with perfect photo\-detectors. In practice, independent Gaussian noise
should be added.

Some key references are the book \citet{leonhardt97}  and the survey papers
\citet{dariano97a, dariano97b,dariano01}.  Of special interest is a maximum
likelihood based approach to the problem that has been taken in recent work by
\citet{banaszeketal00}. We think that it is a major open problem to work out the 
asymptotic theory of this method, taking account of data-driven truncation, and
possibly alleviating the problem of the large parameter-space by using
Bayesian methods. The method should be tuned to the estimation of various
functionals of $\rho$ of interest, and should provide standard errors or
confidence intervals.

\subsection{Quantum Stochastic Processes and Continuous-Time Measurements}\label{ss:qsp}

In this paper we have focussed directly on questions of quantum statistical
inference. Of major related importance are the areas of quantum stochastic
processes and continuous-time quantum measurements. These are currently
undergoing rapid developments, and the concept of quantum instruments,
discussed above, has a key role in parts of this. References to much of this
work are available in \citet{biane95} (see also the more extensive account by
\citep{meyer93}), \citet{percival98}, \citet{holevo01,holevo01book}, 
\citet{belavkin02}, and \citet{barndorffnielsenloubenets02}.

There are quantum analogues of Brownian motion and Poisson processes, and
more generally of L\'{e}vy processes, and a quantum stochastic analysis
based on these. Interesting combinations of classical and quantum stochastic
analysis occur in a variety of contexts, for instance in Monte Carlo simulation studies
of the Markov quantum master equation; \citet{moelmercastin96}
is an important early reference. The Markov quantum master equation is
important particularly in quantum optics which is one of the currently most
active and exciting fields of quantum physics.

Other, mainly mathematically motivated, studies have strongly algebraic
elements, such as in free probability. In this context a variety of
`independence' concepts have turned up, with associated L\'{e}vy processes, etc. 
See \citet{barndorffnielsenthorbjornsen01,barndorffnielsenthorbjornsen02a, 
barndorffnielsenthorbjornsen02b}
and \citet{franzleandreschott01} and references given there. Note, moreover, that
\citet{bianespeicher01} discuss a concept of free Fisher information.

\subsection{Quantum Tomography of Operations}

We have focussed on quantum statistical models where only the state depends on
an unknown parameter. Of great interest is also the situation where an \emph{unknown
operation} acts on a \emph{known} state.

Consider a quantum instrument $\mathcal N$ which produces no data but simply
converts an input state $\rho$ into an output state $\sigma(\rho;\mathcal N)$.
By the general theory, $\sigma(\rho;\mathcal N)=\sum_i n_i\rho n_i^*$ for some collection
of  matrices $n_i$ satisfying $\sum n_i^* n_i=\mathbf 1$. This representation is not unique
but one can fix the $n_i$ by making some identifying restrictions. One could then proceed
to estimate the $n_i$ by feeding the instrument with sufficiently many different 
input states $\rho$, many times, each time carrying out sufficiently many different 
measurements on the output state.

It has recently been discovered that there is an extremely effective short cut to this procedure.
Consider two copies of the original quantum system, supposed here to be of dimension $d$.
Consider the maximally entangled state $\ket\Psi=\sum_j\ket j \otimes \ket j /\sqrt d$ on the
product system.  Now allow the instrument $\mathcal N$ to act on the first component
of the product system, while the second component is left unchanged. The output state,
also on the product system, has density matrix 
$\sigma(\ket\Psi\bra\Psi;\mathcal N\otimes \mathcal I)$ where $\mathcal I$ denotes the identity instrument. It turns out that the output
state completely characterizes $\mathcal N$. In fact, there is a one-to-one correspondence
between on the one hand completely positive data-less instruments $\mathcal N$,
and on the other hand density matrices on the 
product system such that the reduced density matrix of the second component is the 
completely mixed state $\sum_j \ket j\bra j/d$ (i.e., the same as the reduced density matrix of the second component initially, which is left unchanged by the procedure).

Thus one does not need to probe the instrument with many different input states, but can 
effectively probe it with all inputs simultaneously, by exploiting quantum entanglement with
an auxiliary system. The problem has been converted into a quantum statistical model  
$\sigma(\ket\Psi\bra\Psi;\mathcal N\otimes \mathcal I)$  with parameter being the
unknown instrument $\mathcal N$. 

This procedure has been pioneered by \citet{darianolopresti01}
and has already been exploited experimentally.

\subsection{Conclusion}

This paper has, in brief form, presented our present view of a role for
statistical inference in quantum physics. We are keenly aware that many 
relevant parts of quantum physics and quantum stochastics have not
been reviewed, or have only been touched upon.

\paragraph{Acknowledgements}

We gratefully acknowledge Mathematische Forschungsinstitut Oberwolfach 
for support through the Research in Pairs programme, and the European 
Science Foundation's programme on quantum information for supporting a 
working visit to the University of Pavia.

We have benefitted from conversations with many colleagues. 
We are particularly grateful to Elena Loubenets, Hans Maassen, 
Franz Merkl, Klaus M{\o}lmer and 
Philip Stamp.

\bibliographystyle{Chicago}

\bibliography{qiread14lite}

\end{document}